\documentclass[runningheads]{llncs}
\usepackage{amsmath,amssymb,xspace,epsfig,color}
\usepackage[linesnumbered,ruled]{algorithm2e}
\usepackage[tight]{subfigure}
\usepackage{wrapfig}
\usepackage{hyperref}
\usepackage[misc]{ifsym}
\usepackage[T1]{fontenc}
\usepackage{color}

\long\def\commented#1{}
\def\etal{\emph{et~al.}\xspace}

\newcommand{\denselist}{\itemsep 0pt\parsep=1pt\partopsep 0pt}
\newcommand{\bitem}{\begin{itemize}\denselist}
\newcommand{\eitem}{\end{itemize}}
\newcommand{\benum}{\begin{enumerate}\denselist}
\newcommand{\eenum}{\end{enumerate}}

\long\def\commented#1{}
\def\etal{\emph{et~al.}\xspace}

\long\def\soln#1{}

\begin{document}
\title{Network Alignment by \\Discrete Ollivier-Ricci Flow}
%
%
\author{Chien-Chun Ni\inst{1}\orcidID{0000-0002-9082-7401} \and
Yu-Yao Lin\inst{2}\orcidID{0000-0001-9761-5156} \and \\
Jie Gao\inst{3}\Letter\orcidID{0000-0001-5083-6082} \and
Xianfeng Gu\inst{3}\orcidID{0000-0001-8226-5851}
}
\authorrunning{C.-C. Ni et al.}
%
\institute{
Yahoo! Research, Sunnyvale CA 94089, USA \\
\email{cni02@oath.com} \and
Intel Inc., Hillsboro OR 97124, USA \\
\email{yu-yao.lin@intel.com} \and
Stony Brook University, Stony Brook NY 11794, USA \\
\email{\{jgao,gu\}@cs.stonybrook.edu}
}

\maketitle              
\begin{abstract}
In this paper, we consider the problem of approximately aligning/matching two graphs. Given two graphs $G_{1}=(V_{1},E_{1})$ and $G_{2}=(V_{2},E_{2})$, the objective is to map nodes $u, v \in G_1$ to nodes $u',v'\in G_2$ such that when $u, v$ have an edge in $G_1$, very likely their corresponding nodes $u', v'$ in $G_2$ are connected as well.  This problem with subgraph isomorphism as a special case has extra challenges when we consider matching complex networks exhibiting the small world phenomena. In this work, we propose to use `Ricci flow metric', to define the distance between two nodes in a network. This is then used to define similarity of a pair of nodes in two networks respectively, which is the crucial step of network alignment.
Specifically, the Ricci curvature of an edge describes intuitively how well the local neighborhood is connected. The graph Ricci flow uniformizes discrete Ricci curvature and induces a Ricci flow metric that is insensitive to node/edge insertions and deletions. With the new metric, we can map a node in $G_1$ to a node in $G_2$ whose distance vector to only a few preselected landmarks is the most similar. The robustness of the graph metric makes it outperform other methods when tested on various complex graph models and real world network data sets (Emails, Internet, and protein interaction networks)\footnote{The source code of computing Ricci curvature and Ricci flow metric are available: https://github.com/saibalmars/GraphRicciCurvature}. 



\end{abstract}

\section{Introduction}

Given two graphs $G_1$ and $G_2$ with approximately the same graph topology, we want to find the correspondence of their nodes -- node $v$ in $G_1$ is mapped to a node $f(v)$ in $G_2$ such that whenever $u, v$ is connected in $G_1$, $f(u), f(v)$ are likely to be connected in $G_2$ as well. This is called the \emph{network alignment problem} and has been heavily studied~\cite{CONTE:2004cd} with numerous applications including database schema matching~\cite{Fan:2013fm}, protein interaction alignment~\cite{Singh2008-hv,Clark2014-ds,Elmsallati2016-kw,malod2017unified}, ontology matching~\cite{Shvaiko2013-fj}, pattern recognition~\cite{Melnik:2002wd} and social networks~\cite{Zhang2015-ol,Goga2015-ul}.

Network alignment is a hard problem. A special case is the classical problem of \emph{graph isomorphism}, in which we test whether or not two graphs have \emph{exactly} the same topology under a proper correspondence. 
The most recent breakthrough by L\'{a}szl\'{o} Babai~\cite{DBLP:journals/corr/Babai15} provides an algorithm with quasi-polynomial running time $O(\exp{((\log n)^{O(1)})})$, where $n$ is the number of nodes. It still remains open whether the problem is NP-complete or not. For special graphs, polynomial time algorithms have been developed (e.g. trees~\cite{Aho:1974:DAC:578775}, planar graphs~\cite{Hopcroft:1974:LTA:800119.803896}, graphs of bounded valence~\cite{luks1982isomorphism}). 
For general graphs what has been commonly used include heuristic algorithms such as
spectral methods~\cite{Singh2008-hv,Aflalo:2015hd,patro2012global}, random walks~\cite{Tong:2007ut}, optimal cost matching method~\cite{Khan:2013fu} and software packages such as \textsc{Nauty}, \textsc{VF}, and \textsc{VF2}, see~\cite{McKay2014,Cordella:2004fa}. 
The subgraph isomorphism problem, i.e., testing whether one graph is the subgraph of the other, is NP-complete and has been heavily studied as well. 
See the survey~\cite{CONTE:2004cd,Yan:2016:SSR:2911996.2912035,EmmertStreib:2016ft}.

For practical settings, the problem is often formulated as finding a matching in a complete bipartite graph $H=(V_1, V_2, E)$, in which $V_1$ are nodes in $G_1$, $V_2$ are nodes in $G_2$, and edges $E$ carry weights that indicate similarity of the two nodes. Here similarity can be either attribute similarity or structural similarity (quantifying their positions in the network). 
The two graphs are aligned by taking a matching of high similarity in graph $H$. Many algorithms use this approach and they differ by how to define node similarity: L-Graal~\cite{Malod-Dognin2015-do}, Natalie~\cite{El-Kebir2015-mt}, NetAlignMP++~\cite{Bayati2013-mr}, NSD~\cite{kollias2012network}, and IsoRank~\cite{Singh2008-hv}. 
In this paper, our main contribution is to provide a new method to compute \emph{node structural similarity} using the idea of graph curvature and curvature flow. 


\smallskip\noindent\textbf{Our Setting.}
We mainly focus on complex networks that appear in the real world. Consider the Internet backbone graphs captured at two different points in time, we wish to match the nodes of the two graphs in order to understand how the network has evolved over time. Or, consider two social network topologies on the same group of users. It is likely that when two users are connected by a social tie in one network (say LinkedIn), they are also connected in the other network (say Facebook). A good alignment of two networks can be useful for many applications such as feature prediction~\cite{Ewing:2007jq}, link prediction~\cite{Wang:2011fz}, anomaly detection~\cite{Noble:2003cs}, and possibly de-anonymization~\cite{Fu:2015gl,Peng:2014ha}.

We work with the assumption that a small number $k$ (say $2$ or $3$) of nodes, called landmarks, are identified.  Often a small number of landmarks could be discovered by using external information or properties of the networks. For example, in networks with power law degree distributions, there are often nodes with really high degree that can be identified easily.  Given the landmarks, we can find a coordinate for each node $u$ as $[d(u, \ell_1), d(u, \ell_2), \cdots, d(u, \ell_k)]$, where $d(u, \ell_i)$ captures the similarity of $u$ with the $i$th landmark.
This coordinate vector captures the structural position of a node in the network and can be used in the network alignment problem -- two nodes $u_1\in V_1$ and $u_2\in V_2$ can be aligned if their respective coordinates are similar. This approach is motivated by localization in the Euclidean setting, in which three landmarks are used to define the barycentric coordinates of any node in the plane. 
In the complex network setting, this approach faces two major challenges that need to be addressed. 

First, a proper metric that measures the distance from any node to the landmarks in the network is needed. The easiest metric is probably the hop count -- the number of edges on the shortest path connecting two nodes in the network. While the hop count is often used for communication networks, for complex networks it is less helpful as these networks often have a small diameter. 
To look for a better measure, edges shall be properly weighed which is highly application dependent. For the Internet setting~\cite{Lim2005-ft}, measures such as Round-Trip Time (RTT) are used.
In a social network setting, it is natural to weight each edge by its tie strength. 
But these weights are not easy to obtain. Many graph analysis methods use measures that capture local graph structures by common neighborhood, statistics generated by random walks~\cite{Perozzi:2014ib,Grover:2016ex}, etc. 

Second, the approach of using distances to landmarks to locate a node in the network is truly geometric in nature~\cite{fggsz-glrsn-05,kleinberg04triangulation}. But a fundamental question is to decide what underlying metric space shall be used. Low-dimensional Euclidean spaces are often used~\cite{Ham2005-qg,Lafon2006-lb} and spectral embedding or Tutte embedding are popular.
But it is unclear what is the dimensionality of a general complex network and robustness of these embeddings is not fully understood. 

\begin{figure}[tbp]
	\centering
	\subfigure[Ricci Curvature: The Original]{
		\label{fig:karate:subfig:rc}
		\includegraphics[width=0.45\columnwidth]{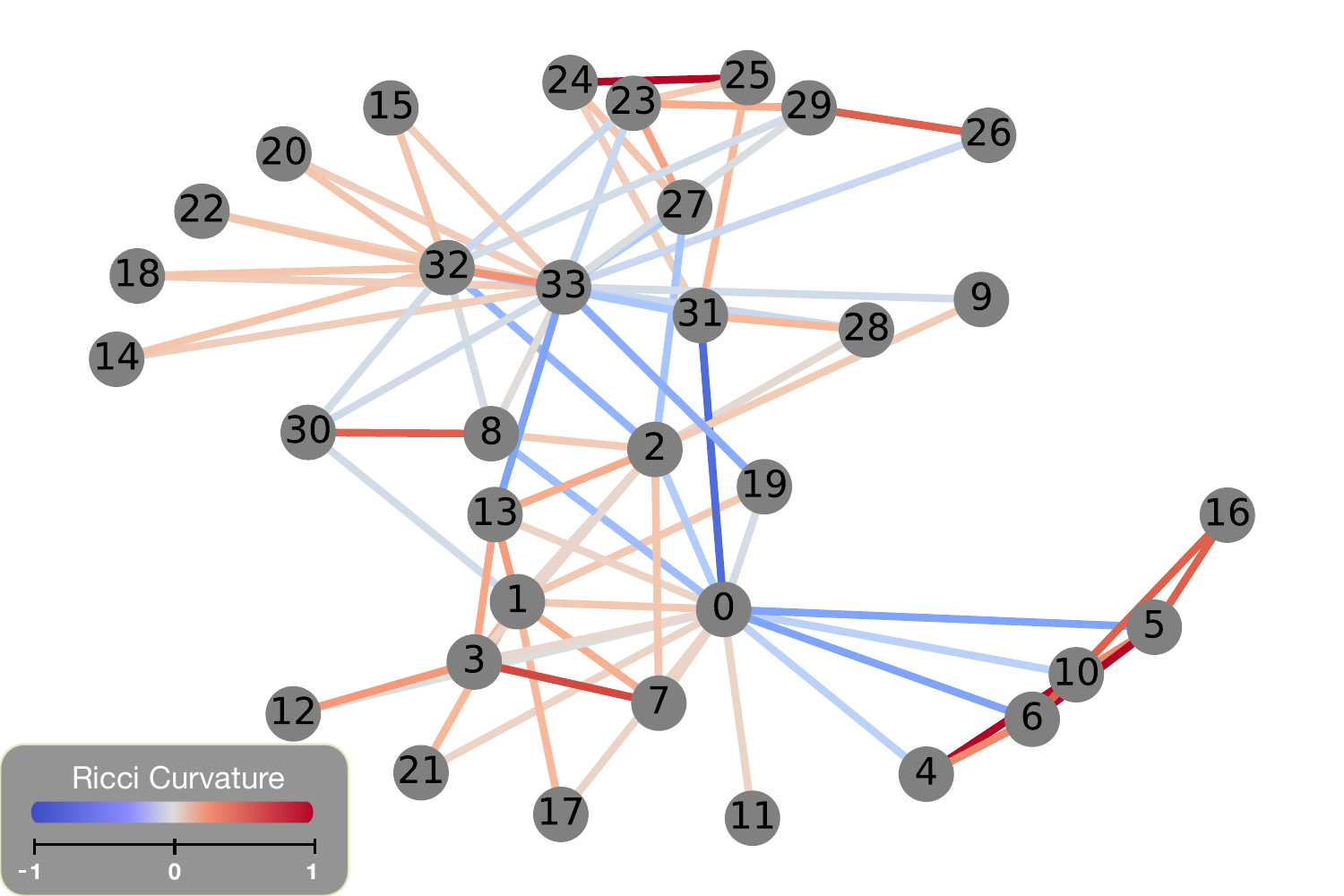}}
	\subfigure[Ricci Curvature: After Ricci Flow]{
		\label{fig:karate:subfig:rf}
		\includegraphics[width=0.45\columnwidth]{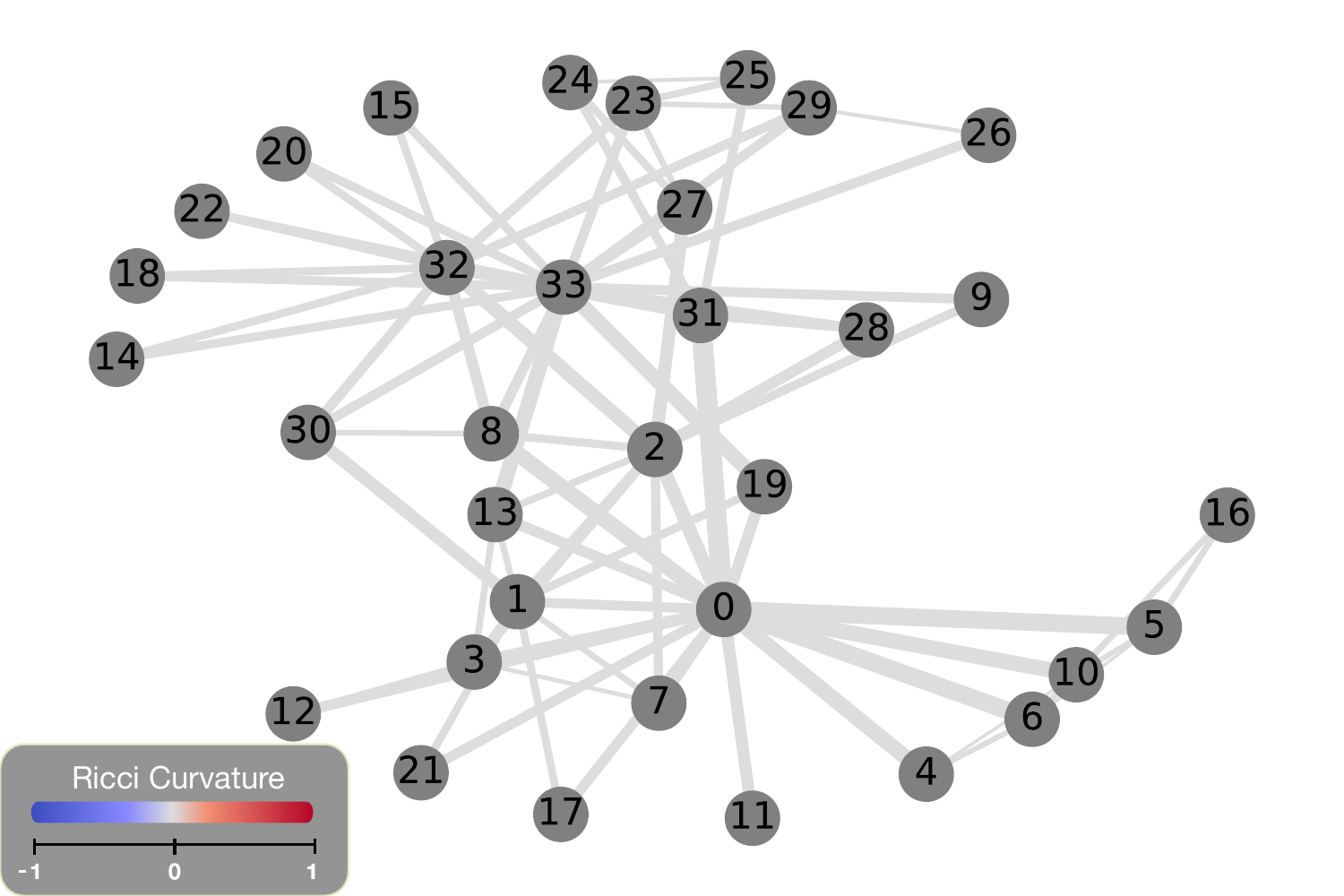}}
	\caption{\scriptsize An example of Ricci curvature on the karate club graph before and after Ricci flow. The colors represent the Ricci curvature while the thickness represents the edge weight. 
    Ricci flow deformed the edge weights until Ricci curvatures converged ($-0.0027$ in this case). 
    }
	\label{fig:karate} 
		
\end{figure}

\smallskip\noindent\textbf{Our Approach.} We address the above problems by using the tool of graph curvature and curvature flow. In the continuous setting of a two dimensional surface, the curvature of a point captures how much it deviates from being flat at the point. The tip of a bump has positive curvature and a saddle point has negative curvature. Curvature flow is a process that deforms the surface (changing the metric) and eventually makes the curvature to be uniform everywhere. 
In our setting, we look at \emph{discrete} curvatures and curvature flow defined on a \emph{graph} and argue that by curvature one can encode and summarize graph structures. 
As will be explained later in more details and rigor, edges that are in a densely connected `community' are positively curved while edges that connect two dense communities are negatively curved. Further, we can define curvature flow (which is an adaptation of surface Ricci flow to the graph setting) such that weights are given to the edges to make all edges have the same curvature. This process in some sense `uniformizes' the network and can be imagined as embedding the network in some intrinsic geometric space. These weights are again intuitive -- edges in a dense community are short while edges that connect two far away communities are long, see Figure~\ref{fig:karate}. We call the shortest path length under such weights as the Ricci flow metric. The Ricci flow metric is designed to improve robustness under insertions and deletions of edges/nodes -- when an edge is removed, the shortest path distance between the two endpoints in the original graph can possibly change a lot and Ricci flow reduces such imbalance. This is useful for matching two graphs that are different topologically.


\begin{figure}[tbp]
	\centering
    
	\subfigure[Spectral (Avg: $7.9\pm 7.2(\%)$)]{
		\label{fig:diff:subfig:sprctral}
		\includegraphics[width=0.45\columnwidth]{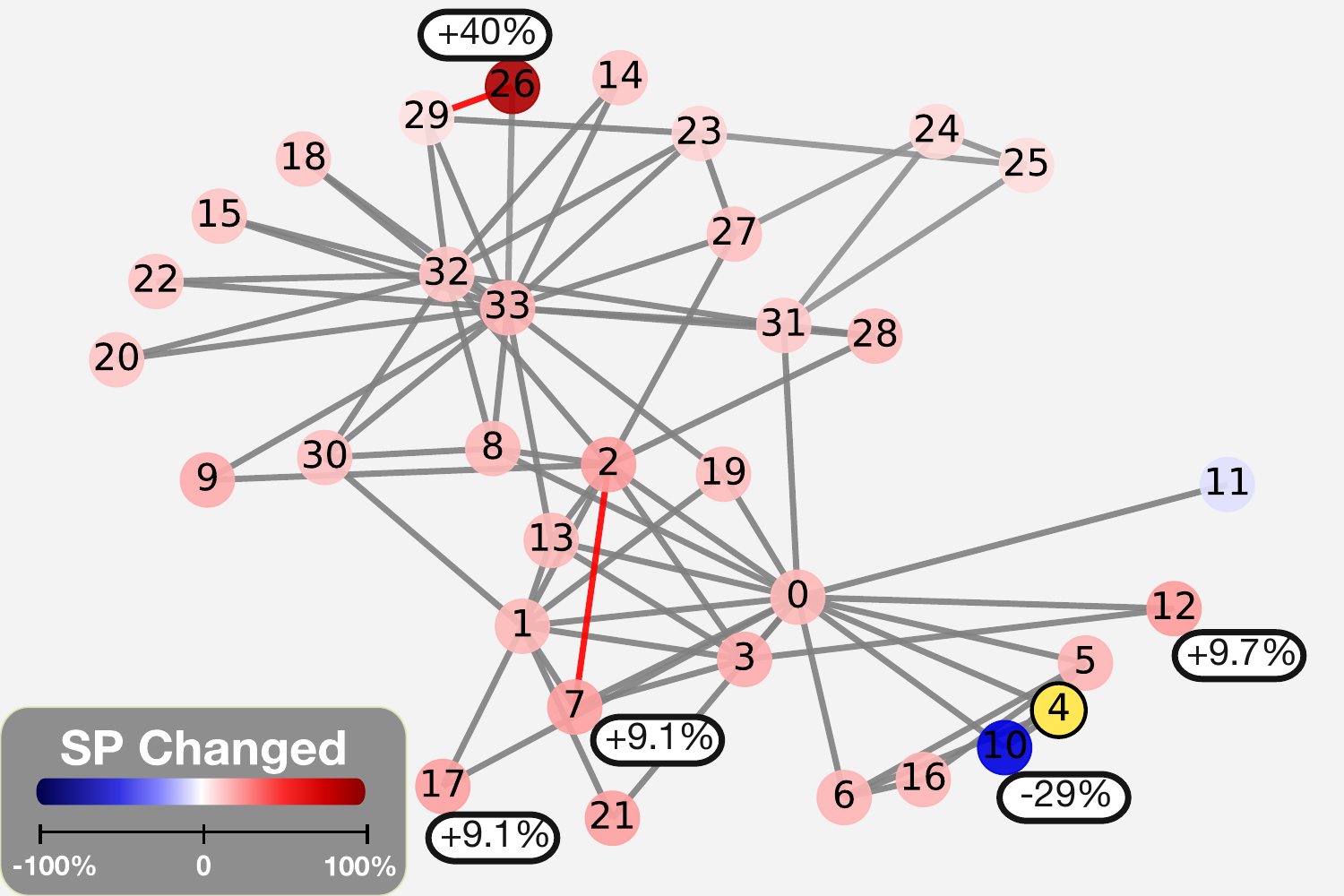}}
	\subfigure[RF Metric (Avg: $1.9\pm 1.1(\%)$)]{
		\label{fig:diff:subfig:urf}
		\includegraphics[width=0.45\columnwidth]{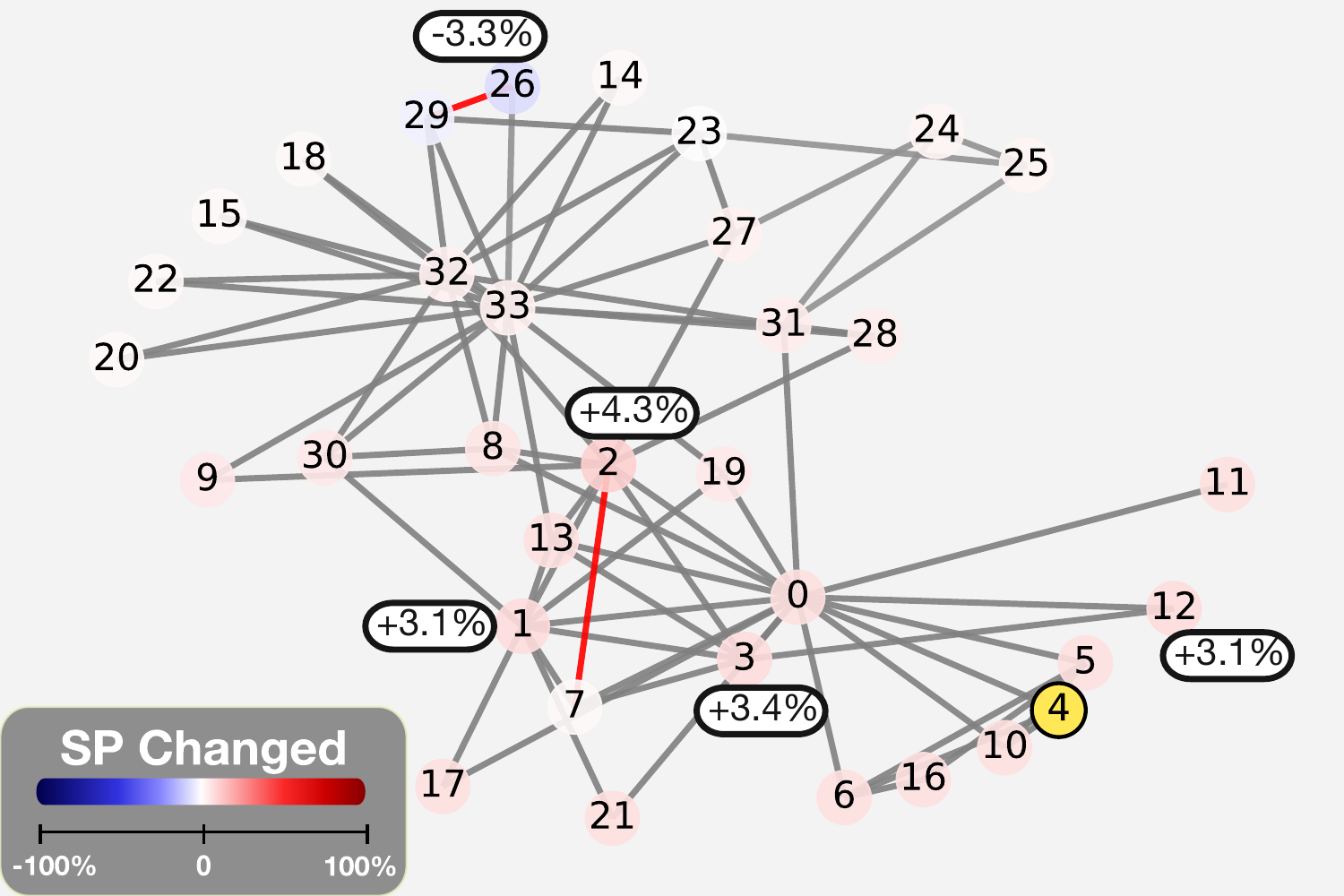}}
	\caption{\scriptsize A comparison of stretch ratios under different metrics started from node $4$ labeled in yellow. Here the stretch ratio defined by the ratio of shortest path changed when nodes or edges are removed from graph. For graph $G$ and $G \setminus \{\overline{v_{2}v_{7}},\overline{v_{26}v_{29}}\}$, the stretch ratios of shortest paths from node $4$ to all other nodes are labeled in color. Here the top five stretches paths are labeled by its ratios next to the nodes. With just two edges removed, the stretch ratios of spring and spectral metric changed drastically while RF metric remains stable.}
	\label{fig:diff} 
		
\end{figure}

Our curvature based coordinates depend only on the network topology and thus are purely structural. But they can also be combined with node attributes (when available) to produce an alignment score.  In our evaluation we have focused on methods that use structural similarities and embeddings such as hop counts, metrics induced by spectral embedding~\cite{Luxburg:2007:TSC:1288822.1288832} or spring/Tutte embedding~\cite{Tutte:1963tz}, and topology based network alignment algorithm  IsoRank~\cite{Singh2008-hv} and NSD~\cite{kollias2012network}. We evaluated our network alignment algorithms on a variety of real world data sets (Internet AS graphs~\cite{Spring2002Rocketfuel}, Email networks~\cite{Guimera:2003dm,Kunegis:2013ih}, protein networks~\cite{Ewing:2007jq,Kunegis:2013ih}) and generated model networks (random regular graphs, Erd\"os-R\'enyi graphs~\cite{erdos59er}, preferential attachment model~\cite{barabasi99emergence} and Kleinberg's small world model~\cite{Kleinberg:2000ug}). 
Experimental evaluations show that the Ricci flow metric greatly outperforms other alternatives.
In particular, most of the embedding methods
perform poorly on random regular graphs due to identical node degree. Methods using hop counts suffer from the problem of not being descriptive, especially when there are only a small number of landmarks. There might be too many nodes with same coordinate under the landmark based coordinates. The spectral embedding and the spring embedding are less robust under edge insertion and deletion. 
IsoRank compares local similarity in a pair of graphs. In a large complex network, it is possible to have nodes that look similar in terms of local structures but globally should not be matched. This may explain why IsoRank does not scale well.


In the following, we first present background knowledge on curvature and Ricci flow~\cite{Hamilton82,tao2008ricci}. We then present graph curvature and curvature flow, how they are used in network alignment, and empirical evaluations.


\section{Background}

\noindent\textbf{Curvature and Ricci Flow on Surfaces.}
Curvature is a measure of the amount by which a geometric object deviates from being flat/straight and has multiple definitions depending on the context. 
Ricci flow was introduced by Richard Hamilton for Riemannian manifolds in 1982 \cite{Hamilton82}. A surface Ricci flow is the process to deform the Riemannian metric of the surface, proportional to the Gaussian curvature, such that the curvature evolves like heat diffusion and becomes uniform at the limit. Intuitively, this behaves as flattening a piece of crumpled paper. 
Surface Ricci flow is a key tool in the proof of the Poincar\'{e} conjecture on 3-manifolds, and has numerous applications in image and shape analysis.  In engineering fields, surface Ricci flow has been broadly applied on a triangulated surface setting for tackling many important problems, such as parameterization in graphics \cite{Jin08TVCGRicci} and deformable surface registration in vision \cite{TPAMI10Ricci}. 

\smallskip\noindent\textbf{Discrete Ricci Curvature.} 
Curvatures for general graphs have only been studied over the past few years~\cite{Ollivier:2009kh,Lin:2011tt,Chung:1996ey,Bakry:1985gp,Lott:2009fk,Sturm:2006wy,Bonciocat:2014ul}. 
The definitions of curvatures that are easier to generalize to a discrete graph setting are sectional curvature and Ricci curvature. Consider a point $x$ on a surface $M$ and a tangent vector $v$ at $x$ whose endpoint is $y$. Take another tangent vector $w_x$ at $x$ and imagine transporting $w_x$ along vector $v$ to be a tangent vector $w_y$ at $y$. Denote the endpoints of $w_x, w_y$ as $x', y'$. If the surface is flat, then $x, y, x', y'$ would constitute a parallelogram. Otherwise, the distance between $x', y'$ differs from $|v|$. The difference can be used to define sectional curvature~\cite{Ollivier:2009kh}. 


Sectional curvature depends on two tangent vectors $v, w$. Averaging sectional curvature over all directions $w$ gives the Ricci curvature which only depends on $v$. Intuitively, if we think of a direction $w$ at $x$ as a point on a small ball $S_x$ centered at $x$, on average Ricci curvature controls whether the distance between a point of $S_x$ and the corresponding point of $S_y$ is smaller or larger than the distance $d(x, y)$. To allow for a general study, Ollivier~\cite{Ollivier:2009kh} defined a Ricci curvature by using a probability measure $m_x$ to represent the ball $S_x$. 
Later Ollivier Ricci curvature has been applied in different fields, for distinguishing between cancer-related genes and normal genes~\cite{Sandhu:2015fm}, for understanding phylogenetic trees~\cite{Whidden:2015hi}, and for detecting network features such as backbone and congestions~\cite{Ni:2015jt,Wang:2014fi,Wang:2016hf,narayan2011large,kennedy2013hyperbolicity}.
There was very limited amount of theory on discrete Ollivier-Ricci curvature on graphs, and nearly none on graph Ricci flow. 
In the future work section of~\cite{Ollivier:2010gg} Ollivier suggested that people should study the discrete Ricci flow in a metric space $(X,d)$ by evolving the distance $d(x,y)$ on $X$ according to the Ricci curvature $\kappa(x,y)$ between two points $x,y\in X$: $$\frac{d}{dt}d(x,y) = -\kappa(x,y)d(x,y).$$ We are the first to study Ollivier-Ricci flow in the graph setting. 
We also suggest variants of the Ollivier-Ricci curvature that admits much faster computation and empirically evaluate properties of the Ricci flow metrics and applications in network alignment.

\section{Theory and Algorithms} 
\label{sec:preliminary}
\noindent\textbf{Ricci Curvature on Graphs} 
\def\Deg{\operatorname{Deg}}
For an undirected graph $G=(V,E)$, the Ollivier-Ricci curvature of an edge $\overline{xy}$ is defined as follows. Let $\pi_x$ denote the neighborhood of a node $x\in V$ and $\Deg(x)$ is the degree of $x$. For a parameter $\alpha\in [0,1]$, define a probability measure $m_{x}^{\alpha}$:
\begin{equation*}
m_{x}^{\alpha}(x_i)=
	\begin{cases}
	\alpha & \mbox{\ if } x_i = x\\
	\left( 1-\alpha\right)/\Deg(x) & \mbox{\ if } x_i \in \pi_x\\
	0 & \mbox{\ otherwise },
	\end{cases}
\end{equation*}
Suppose $w(x, y)$ is the weight of edge $\overline{xy}$ and $d(x, y)$ the shortest path length between $x$ and $y$ in the weighted graph. 
The \emph{optimal transportation distance} (OTD)  between $m_{x}^{\alpha}$ and $m_{y}^{\alpha}$ is defined as the best way of transporting the mass distribution $m_{x}^{\alpha}$ to the mass distribution $m_{y}^{\alpha}$:
\begin{equation*}
W(m_{x}^{\alpha}, m_{y}^{\alpha}) = \inf\limits_{M}\sum_{x_{i},y_{j}\in V}d(x_{i},y_{j})M(x_{i},y_{j})
\end{equation*}
where $M(x_{i},y_{j})$ is the amount of mass moved from $x_i$ to $y_j$ along the shortest path (of length $d(x_i, y_j)$) and we would like to take the best possible assignment (transport plan) that minimizes the total transport distance.
The discrete Ollivier-Ricci curvature~\cite{Ollivier:2009kh} is defined as follows
\begin{equation*}
\kappa^{w}(x,y)=1 - \dfrac{W(m_{x}^{\alpha}, m_{y}^{\alpha})}{d(x,y)}.
\end{equation*}
In this paper, $\kappa^{w}(x,y)$ is called the OTD-Ricci curvature.

The OTD-Ricci curvature describes the connectivity in the local neighborhood of $\overline{xy}$~\cite{Ni:2015jt}. If  $\overline{xy}$ is a bridge so the nodes in $\pi_x$ have to travel through the edge  $\overline{xy}$ to get to nodes in $\pi_y$,  $W(m_{x}^{\alpha}, m_{y}^{\alpha})> d(x, y)$ and the curvature of $\overline{xy}$ is negative. Similarly, if the neighbors of $x$ and the neighbors of $y$ are well connected (such that $W(m_{x}^{\alpha}, m_{y}^{\alpha})< d(x, y)$), the curvature on $\overline{xy}$ is positive.


The optimal transportation distance can be solved by linear programming (LP) to find the best values for $M(x_{i},y_{j})$:  
\begin{equation*}
	\begin{aligned}
		\mbox{Minimize:} \quad& \sum_{i,j} d(x_i,y_j)M(x_{i},y_{j}) \\
		\mbox{s.t.}\quad& \sum_j M(x_{i},y_{j})= m_x ^{\alpha}(x_i), \forall i \quad 
		\mbox{and} \quad \sum_i M(x_{i},y_{j})=m_{y}^{\alpha}(y_j),  \forall j\\
	\end{aligned}
\end{equation*}

The computation of the linear program on large networks may be time-consuming. 
To address the computational challenges, we define a variant of the OTD-Ricci curvature by using a specific transportation plan instead of the optimal transportation plan. We take the \textbf{average transportation distance} (ATD) $A(m_{x}^{\alpha}, m_{y}^{\alpha})$, in which we transfer an equal amount of mass from each neighbor $x_i$ of $x$ to each neighbor $y_j$ of $y$ and transfers the mass of $x$ to $y$. 
One could easily verify that this is a valid transportation plan.
Thus, the discrete Ricci curvature by the average transportation distance (ATD) is defined as:
\begin{equation*}
\kappa^{a}(x,y)=1 - \dfrac{A(m_{x}^{\alpha}, m_{y}^{\alpha})}{d(x,y)},
\end{equation*}
Since we remove the LP step in the computation, the computational complexity of the ATD-Ricci curvature is drastically improved. As will be presented later, our experimental results show that computing discrete Ricci flow using the ATD-Ricci curvature maintains and even enhances the robustness of graph alignment. 
In this paper, we fix $\alpha=0.5$ and simplify the notation of discrete Ricci curvature by $\kappa(x,y)$ in the discussion of discrete Ricci flow.


\smallskip\noindent\textbf{Discrete Ricci Flow} 
Ricci flow is a process that deforms the metric while the Ricci curvature evolves to be uniform everywhere. 
For any pair of adjacent nodes $x$ and $y$ on a graph $G=(V,E)$, we adjust the edge weight of $\overline{xy}$, $w(x, y)$, by the curvature $\kappa(x,y)$:
$$ w_{i+1}(x,y)=w_{i}(x,y)-\epsilon \cdot \kappa_{i}(x,y) \cdot w_{i}(x,y) ,\quad \forall \overline{xy} \in E,$$
where $\kappa_{i}(x,y)$ is computed using the current edge weight $w_{i}(x,y)$. The step size is controlled by $\epsilon > 0$ and we take $\epsilon = 1$ in our experiment.

After each iteration we rescale the edge weights so the total edge weight in the graph remains the same, since only relative distances between nodes matter in a graph metric. A pseudo-code is presented in the appendix. 

\smallskip\noindent\textbf{Ricci Flow Metric} 
When graph Ricci flow converges, each edge $\overline{xy}$ is given a weight $w(x, y)$. We denote the shortest path metric with such weights to be the \emph{Ricci Flow Metric}, denoted as $d(x, y)$. To understand the metric, notice that when the Ricci flow converges the following is true. Here we take step size $\epsilon=1$:
$$ (w(x,y)-\kappa(x,y) \cdot  w(x,y))\cdot N\approx w(x,y)$$
where $$N = \frac{|E|}{\sum_{\overline{xy} \in E} (w(x,y)-\kappa(x, y)w(x, y))}.$$
Denote the transportation distance between the two probability measures $m_{x}^{\alpha}$ and $m_{y}^{\alpha}$ to be $T(x,y)$, we have
$$\kappa(x,y)\approx 1-\frac{T(x,y)}{d(x,y)}, \quad \frac{T(x,y)}{d(x,y)}\approx \frac{1}{N}.$$


To understand what this means, recall that $T(x, y)$ represents the distances from $x$'s neighborhood to $y$'s neighborhood. Before we run Ricci flow, this value for different edges can vary a lot -- in the neighborhood of positively curved edges there are many `shortcuts' making $T(x, y)$ to be significantly shorter than $d(x, y)$, while in the neighborhood of negatively curved edges $T(x, y)$ is longer than $d(x, y)$. The purpose of Ricci flow is to re-adjust the edge weights to reduce such imbalance. Suppose we remove an edge $\overline{xy}$ with negative curvature before Ricci flow and break the shortest path from $x'$ (a neighbor of $x$) to $y'$ (a neighbor of $y$). The alternative path from $x'$ to $y'$ in the neighborhood of $\overline{xy}$ tends to become much longer, as $T(x, y)/d(x, y)$ is large. Thus, the change to the shortest path metric is significant. However, after the Ricci flow, with the new edge weights, the alternative path from $x'$ to $y'$ in the neighborhood of $\overline{xy}$ may still get longer but not as long.
Thus, the change in the shortest path length is less significant. 
The Ricci flow metric is more robust when edges and nodes are randomly removed. 
To capture this property, we define the uniformity of graph metric using the variation of $T(x,y)/d(x,y)$ for different edges $\overline{xy}$. 

\begin{definition}[Metric Uniformity]
	Given a graph $G=(V,E)$ with edge weight $w(x,y)$ on $\overline{xy}\in E$,
the metric uniformity is defined by the interquartile range (IQR), i.e., of $T(x,y)/d(x,y)$ over all edges, the difference between $75th$ and $25th$ percentiles.
\end{definition}

The metric uniformity measures the diversity of $T(x,y)/d(x,y)$ over all edges around the median. Lower metric uniformity indicates that $T(x,y)/d(x,y)$ are less dispersed about the median. Therefore, the corresponding metric is more robust upon node/edge insertions and deletions.


\section{Network Alignment by Ricci Flow Metric} 
\label{sec:graph_matching_by_discrete_ricci_flow}

Given two graphs $G_1$ and $G_2$, suppose there are $k$ landmarks $L=\{\ell_i | i=1, 2, \cdots, k\}$ in both $G_1$ and $G_2$ with known correspondence, $k$ is a small constant such as $3$ or $4$. 
These landmarks may be known beforehand by external knowledge. 
We use these landmarks to find the correspondence of nodes in $G_1$ and $G_2$. Specifically, we represent each node $v\in G_j$ by its relative positions to the landmarks  $v_L=[d_j(v, \ell_1), d_j(v, \ell_2), \cdots, d_j(v, \ell_k)]$, where $d_j(v, \ell_i)$ denotes the shortest distance from $v$ to $\ell_i$ in graph $G_j$ using the Ricci flow metric. 
We define the cost of matching $u\in G_1$ with $v\in G_2$ by the 2-norm of the difference between $u_L$ and $v_L$,
$$C_{uv}=||u_L-v_L||_2.$$
The smaller $C_{uv}$ is, the more similar $u, v$ are. The alignment problem can be formulated as finding a low-cost matching in the complete bipartite graph $H=(V_1 \cup V_2, E)$ where the edge in $E$ connecting $u\in V_1$ and $v\in V_2$ has weight $C_{uv}$.

\smallskip\noindent\textbf{Matching Algorithms.}
To find a low cost matching in the bipartite graph or a similarity matrix, we can apply the following two algorithms.
\begin{itemize}
	\item The Hungarian min-cost matching algorithm~\cite{hopcroft1973n} finds a matching which minimizes $\sum_{u\in G_{1},v\in G_{2}}C_{uv}$ in $O(|V|^{3})$ time. 
	\item A greedy matching method iteratively locates the minimal $C_{uv}$, records the node pair $(u,v)$, and removes all elements involving either $u$ or $v$, until the nodes of either $G_1$ or $G_2$ are all paired. 
\end{itemize}
We note that in prior network alignment algorithms, such as IsoRank and NSD, greedy matching was unanimously selected due to its efficiency. We tested both algorithms with the results on Hungarian algorithms presented in the appendix. 

\smallskip\noindent\textbf{Matching Accuracy.} 
To evaluate the accuracy of the matching results, one idea is to count how many nodes are correctly matched with respect to their IDs in the ground truth. 
But this measure may be too strict. For example, if $G_1$ and $G_2$ are both complete graphs of the same size, any matching of nodes in $G_1$ and nodes in $G_2$ should be considered to be correct. It is well known that social networks have community structures and many different levels of node equivalence and symmetry. There are different definitions for two nodes $u, v$ in a graph $G$ to be equivalent: 
\bitem 
\item \emph{Structural equivalence}: $u$ and $v$ have exactly the same set of neighbors; 
\item \emph{Automorphic equivalence}: if we relabel the nodes in an automorphic transformation (i.e., the graph after relabeling is isomorphic to the original graph), 
the labels of $u, v$ are exchanged; 
\item \emph{Regular equivalence~\cite{Lorrain1971-ev}}: $u, v$ are connected if they are equally related to equivalent others. That is, regular equivalence sets are composed of nodes who have similar relations to members of other regular equivalence sets. 
\eitem 
Regular equivalence is the least restrictive of the three definitions of equivalence, but probably the most important for the sociologists as it captures the sociological concept of a ``role''. 
Previous quantitative measures of regular equivalence~\cite{leicht2006vertex} unfortunately, are not very effective in identifying the global symmetric structures. For example, the two terminal nodes of a path graph cannot be classified to the same regular equivalence class. 

\smallskip\noindent\textbf{Connected Equivalence.} 
Motivated by regular equivalence, we would like to consider two nodes to be \emph{connected equivalence} if they have similar connections to other nodes. Specifically, we compute the length of shortest paths (using the Ricci flow metric) from $u$ and $v$ to all other nodes, except $u, v$ themselves, in a fixed order in $G_1$ or $G_2$, and denote the results by two vectors $u_L$ and $v_L$. If the two vectors are similar (i.e., $||u_{L}-v_{L}||_2 < \epsilon$ for some small $\epsilon>0$), then we say that $u, v$ are equivalent and the matching of $u$ to $v$ is correct.
\section{Evaluation}

In this section, we demonstrate the performance of Ricci flow metric 
and its power on network alignment. Since Ricci flow metric only used topology feature, we compare RF-OTD, RF-ATD with two topology based network alignment algorithms IsoRank~\cite{Singh2008-hv} and NSD~\cite{kollias2012network} (other network alignment algorithms require node label/attributes), and three different embedding metrics: spectral embedding, spring embedding, and hop count. We evaluated the performance of these algorithms in both model graphs and real world data. 
The descriptions of these methods as well as the data sets are presented in Appendix~\ref{sub:data_sets} and Appendix~\ref{sub:matching_evaluations}.
Our main observations are as follows:

\begin{enumerate}
    \item Ricci flow metrics (both RF-OTD and RF-ATD) are much more robust against edge/node insertion and deletions compared to others on noisy graph alignment problem. 
    \item Ricci flow metrics greatly outperform previous topological based network alignment algorithms (IsoRank, NSD) as well as other metrics and similarity measures on noisy graph alignment problem. In many experiments our algorithm achieves more than $90\%$ matching accuracy while other methods achieve accuracy below $30\%$.
	\item While greedy matching performs well in model graphs, min-cost matching (i.e, Hungarian algorithm) is more suitable for real world graph.
\end{enumerate}

Due to space limit, please refer to Appendix~\ref{sec:eval_result} for further discussion. 


\smallskip\noindent\textbf{Experiment Setup.} Given a 
graph $G_1$, we remove $n$ nodes or edges (less than $1\%$ ) uniform randomly to create $G_2$ as a noisy graph. We then perform the alignment of $G_1$ and $G_{2}$. 
Notice that in this case, we can also regard $G_2$ as a graph with random noises added to $G_1$. Therefore, we only present the results of node deletion cases.

The graph alignment problem is solved in two steps. First we construct a similarity matrix that records the cost of matching node $i$ in $G_1$ with node $j$ in $G_2$ for all possible $i, j$; then we perform Hungarian algorithm or greedy matching to match nodes in $G_1$ with nodes in $G_{2}$. We evaluate the algorithm performance by using connected equivalence. 

For IsoRank and NSD, the similarity matrix is the output of the algorithm; for landmark based method, the similarity matrix is defined by $\ell_2$ norm of their landmark distance vectors. Here the distance metric is defined by RF-OTD and RF-ATD\footnote{With 50 Ricci flow iterations, $\epsilon=1$, $\alpha=0.5$}, distances induced by spectral embedding (dim=$2$), spring embedding, and hop count respectively. The performance of each metric is evaluated by varying the number of removed nodes/edges $n$ and the number of landmarks $k$. To eliminate the dependency of landmark selection, the matching accuracy is averaged by $10$ experiments on different sets of landmarks. These landmarks are chosen such that every newly added landmark is furthest away from the landmarks chosen so far.

\begin{figure}[tbp]
	\centering
	\subfigure[Metric uniformity comparison]{
		\label{fig:edge-compare:subfig:edge}
		\includegraphics[width=0.45\columnwidth]{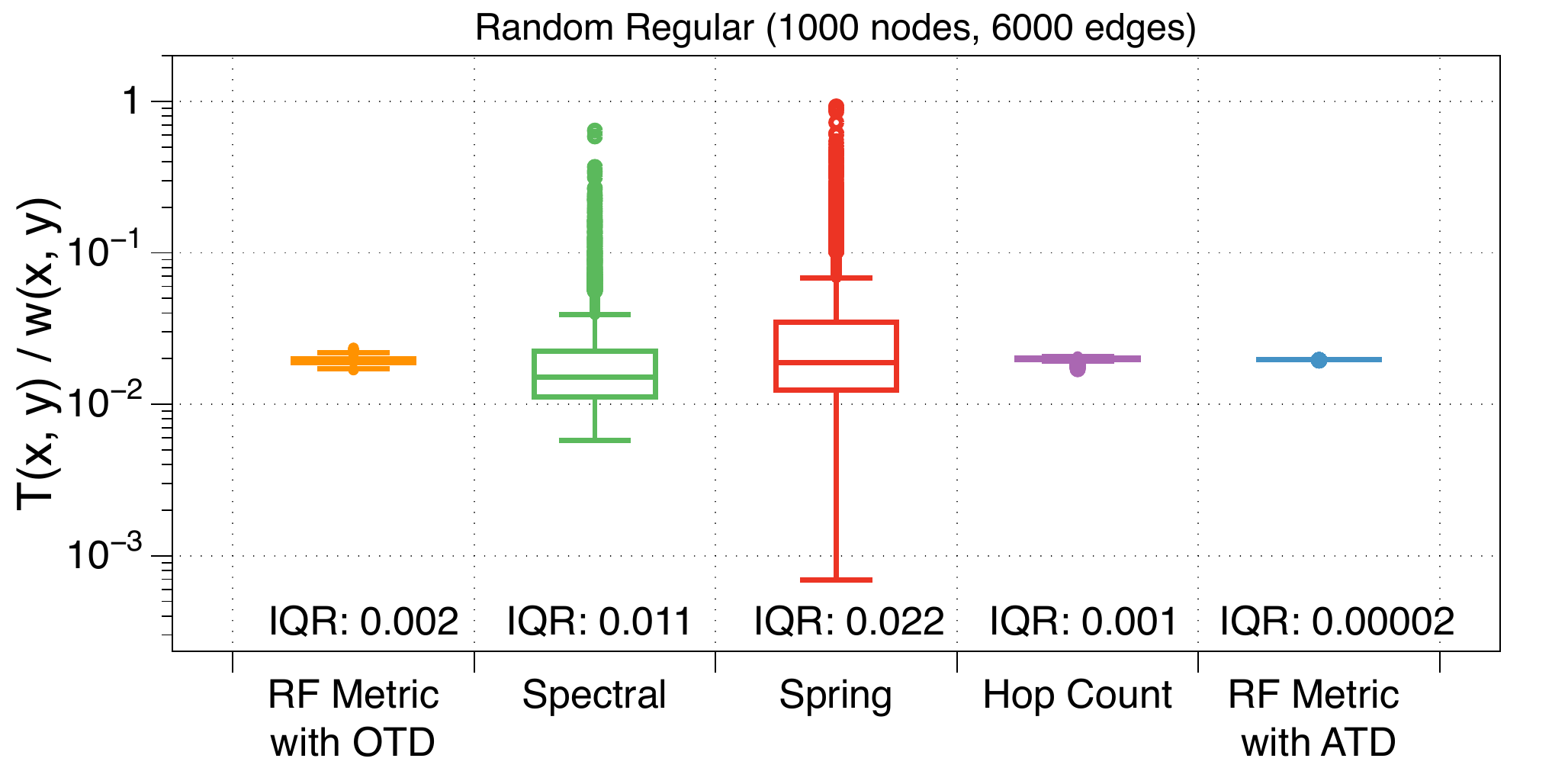}}
	\subfigure[Stretch ratios comparison]{
		\label{fig:edge-compare:subfig:stretch}
		\includegraphics[width=0.45\columnwidth]{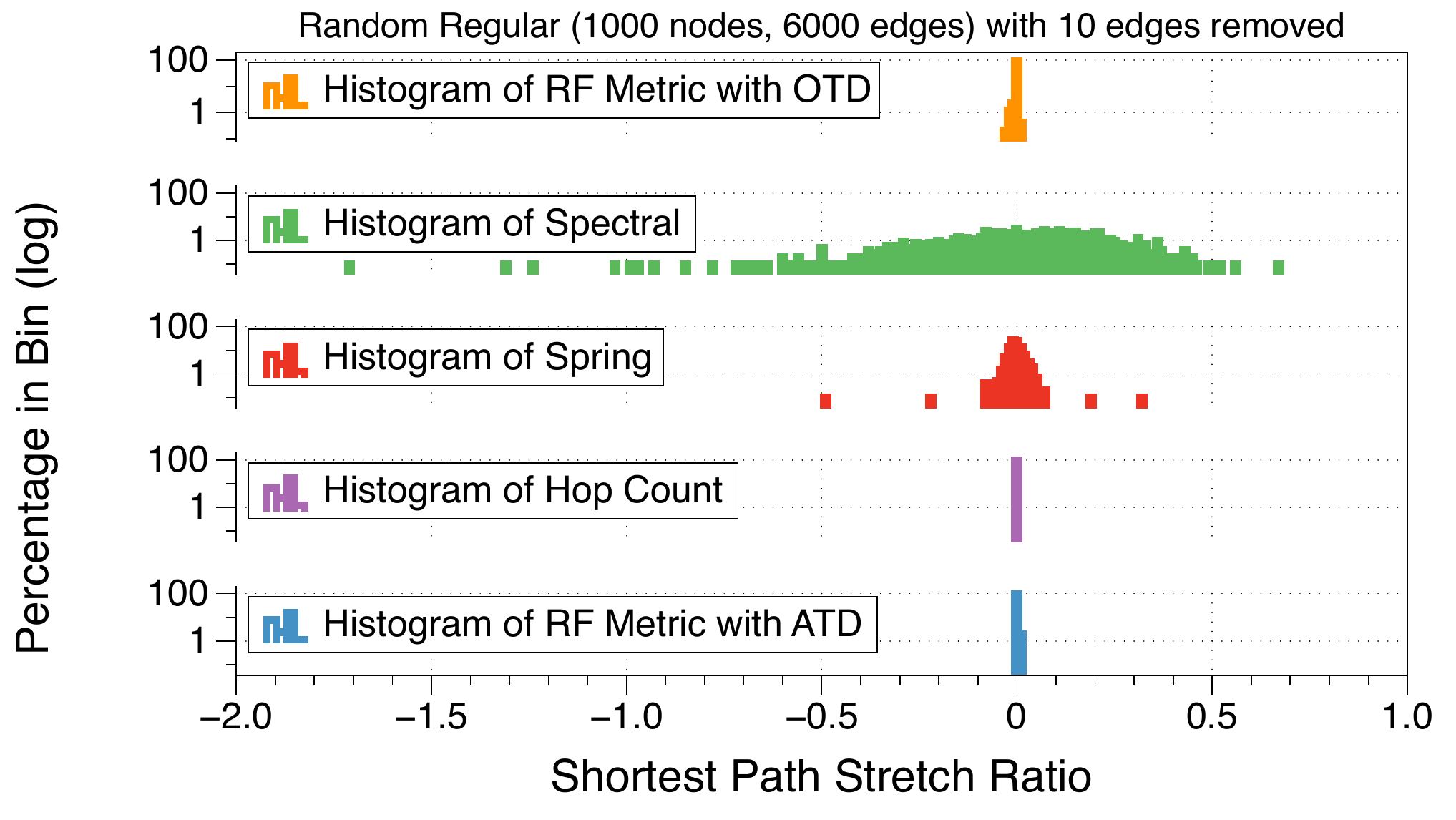}}
	\caption{\scriptsize Figure~\ref{fig:edge-compare:subfig:edge} shows metric uniformity comparison of all metrics. The box plot is for $T(x,y)/w(x,y)$ with respect to each metric. We define the IQR (InterQuartile Range) of the box plot to be the metric uniformity. The smaller the metric uniformity is, the more robust the metric is under random edge removals. Figure\ref{fig:edge-compare:subfig:stretch} demonstrates a comparison of the stretch ratios of shortest path length over all methods.}
	\label{fig:edge-compare} 
		
\end{figure}

\smallskip\noindent\textbf{Metric Uniformity} 
We first analyze the metric uniformity over all metrics on a random regular graph with $1000$ nodes and $6000$ edges. The box plot result is illustrated in Figure~\ref{fig:edge-compare:subfig:edge}. RF-ATD yields the best metric uniformity with the smallest IQR of $2e-5$. RF-OTD also performs well with IQR as $0.002$. Spectral and spring embedding behave poorly. These performances are directly related to the accuracy in graph alignment. The metric with low metric uniformity is more stable when nodes/edges are missing, see Figure~\ref{fig:diff}. Notice that hop count metric generates a small IQR=$0.001$. This is because there are often multiple shortest paths (in terms of hop count) connecting two nodes in the network so hop count is actually a fairly robust measure. But using it for graph alignment is still limited by its lack of descriptive power.

For further analysis that metric uniformity indeed captures the robustness of shortest path metric,  we compute the stretch of the shortest path length between a pair of nodes in a random regular graph when $10$ edges are randomly removed, shown in Figure~\ref{fig:edge-compare:subfig:stretch}. Consider two nodes $u,v$ in both $G_1$ and $G_2$, we denote the length of the shortest path from $u$ to $v$ by $d_{G_{i}}^m(u,v)$ under a metric $m$, where $i=1,2$. We define the stretch ratio of $G_1$ and $G_2$ as $s(u,v)=(d_{G_1}^m(u,v)-d_{G_2}^m(u,v))/d_{G_1}^m(u,v)$. 
The stretch ratio captures the changes of the shortest path length. A larger stretch ratio means the shortest path length changes more. In Figure~\ref{fig:edge-compare:subfig:stretch}, we collect the stretch ratios from one random node to all of the other nodes based on different metrics, and plot the distribution of these stretch ratios as a histogram. It turns out that hop count, Ricci flow metric with OTD and with ATD result in smaller stretch ratios, while spectral embedding is the most vulnerable one with edge deletions.

\begin{figure}[bp]
    \centering
    \includegraphics[width=0.5\columnwidth]{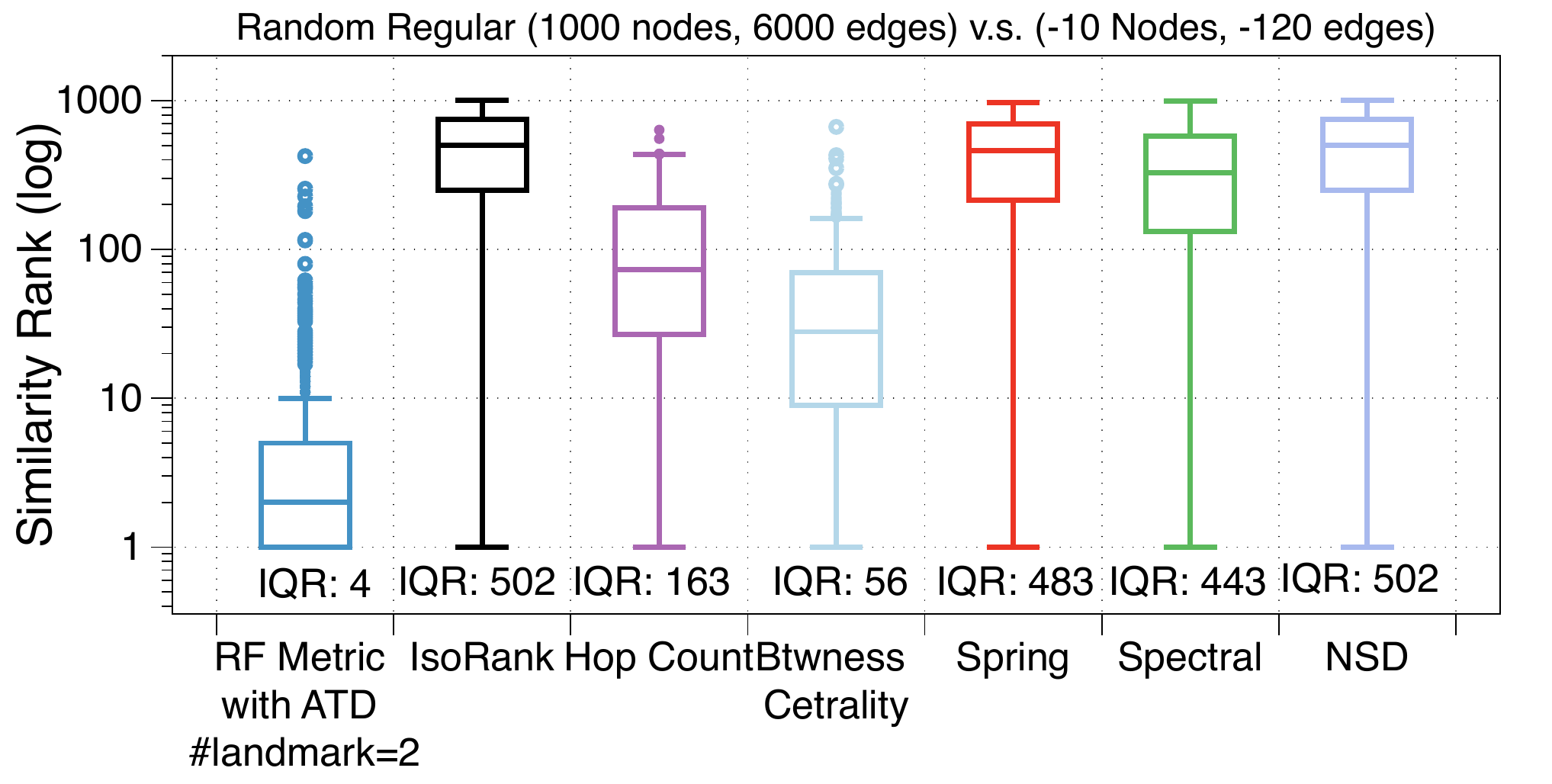}
    \caption{\scriptsize The comparison of the similarity rank of all methods. The similarity matrix represents the pairwise similarity between a random regular graph with $1000$ nodes and $6000$ edges and the previous graph with $10$ random nodes and $120$ corresponding edges removed.}
    \label{fig:sim_matrix}
    
\end{figure}


\smallskip\noindent\textbf{Similarity Matrix} 
Here we test the similarity matrix of every pairwise node similarity of two graphs $G_1$, $G_2$. $G_1$ is a random regular graph with $1000$ nodes and $6000$ edges and $G_2$ removes $10$ nodes randomly from $G_1$. 
We check the performance of the similarity matrix as follows. For every node $n_1 \in G_1$ and the corresponding node $n_2$ in $G_2$ in the ground truth, we check the rank of $n_2$ in the sorted list ranked by similarity values with $n_1$. A good similarity matrix should rank $n_2$ as the most similar one. A lower ranking indicates better performance of the similarity matrix. 
We show the results in Figure~\ref{fig:sim_matrix} with $2$ landmarks. 
Thanks to the metric uniformity, our method yields the best performance with average similarity rank of $2$ while other methods are at least $10$ times higher. 



\begin{figure}[tbp]
	\centering
	\subfigure[Random Regular Graph ]{
		\label{fig:matching:subfig:model}
		\includegraphics[width=0.45\columnwidth]{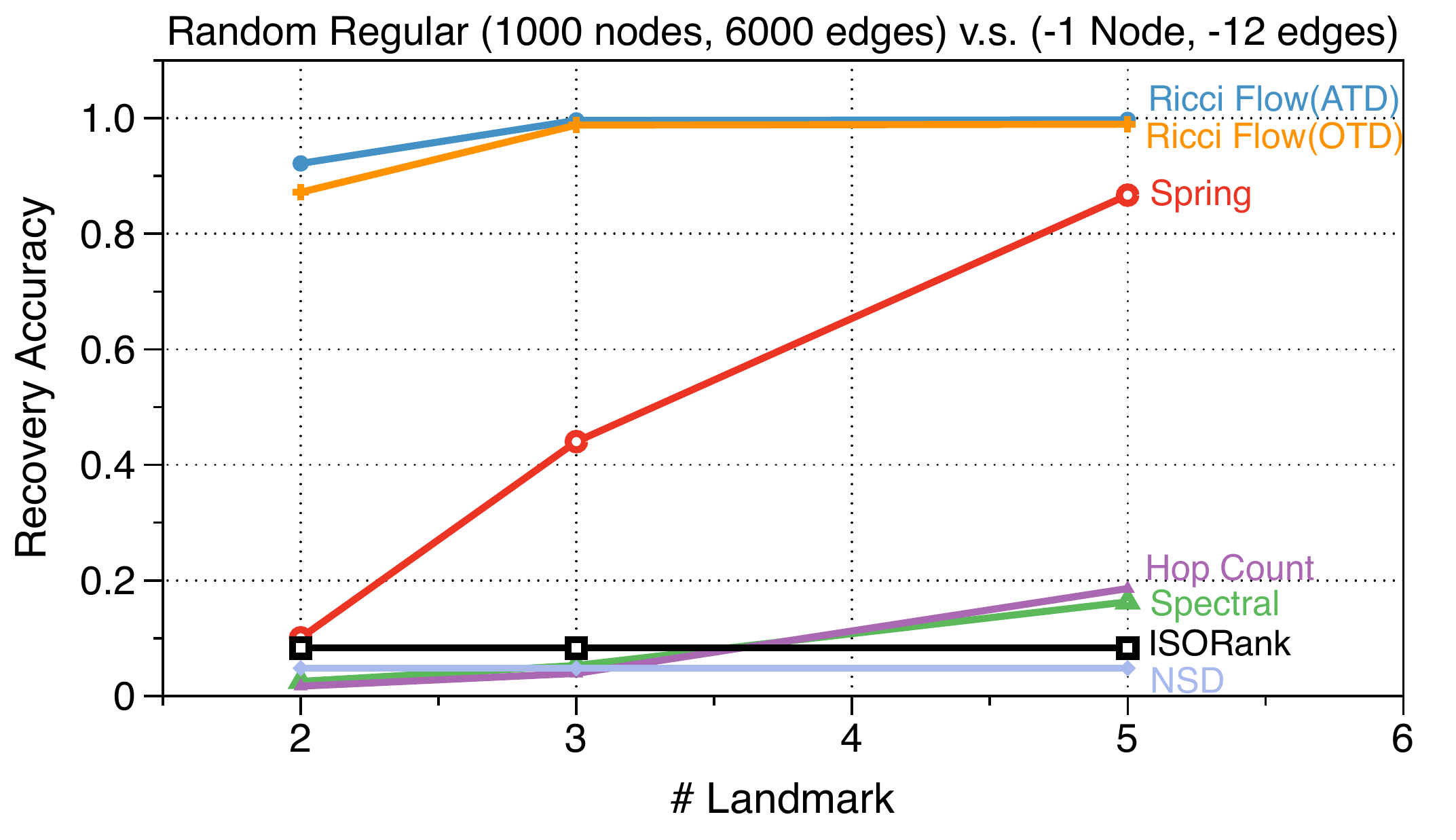}}
	\subfigure[Protein-Protein Interaction]{
		\label{fig:matching:subfig:real}
		\includegraphics[width=0.45\columnwidth]{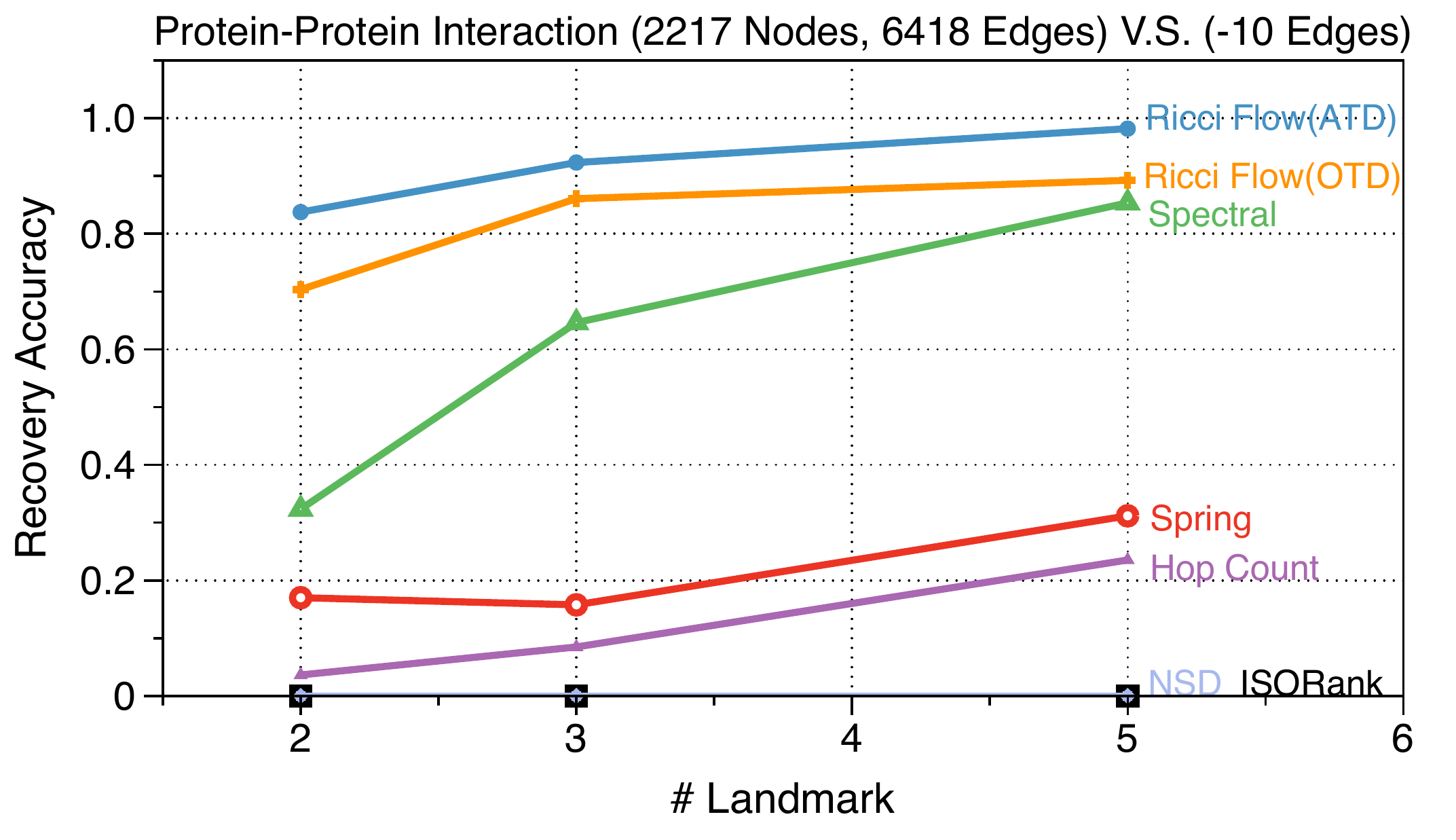}}
	\caption{\scriptsize A comparison of noisy graph alignment results by Hungarian algorithm and connected equivalence on a random regular graph(Figure~\ref{fig:matching:subfig:model}) and protein-protein interaction graph(Figure~\ref{fig:matching:subfig:real}).}
	\label{fig:matching} 
		
\end{figure}

\smallskip\noindent\textbf{Network Alignment} 
Here we demonstrate the matching accuracy results on noisy graph alignment problem for a random regular graph ($1000$ nodes and degree $12$ as $G_1$, and $G_2$ by randomly removing $1$ node and $12$ edges from $G_1$) and a real word protein-protein interaction graph (to be aligned with a graph of $10$ edges randomly removed). 
Figure~\ref{fig:matching} shows that while most of the methods failed to align the graph correctly, the Ricci flow metric performs well with only $2$ landmarks. Here since IsoRank and NSD method do not require landmarks, the performance over different landmark is shown as a straight line. The result shows that RF-ATD, which is computationally much more efficient, performs equally well as RF-OTD. Notice that the methods with poor metric uniformity as shown in Figure~\ref{fig:edge-compare} also result in poor performance here. This supports the importance of metric uniformity in network alignment. More evaluation on different model networks and real networks also support this claim. The performance of alignment accuracy also effected by the portion of nodes and edges added/removed, the noisy graph alignment problem become harder with more noise. 
Please see Appendix~\ref{sub:matching_evaluations} for further details.


\section{Conclusion} 
\label{sec:conclusion}
In this paper, we have presented a framework to endow a graph with a novel metric through the notion of discrete Ricci flow with an application to network alignment. 
From the experimental results, we found that 1) the graph Ricci curvature converges through the discrete Ricci flow and; 2) Ricci flow metric on a graph is fairly stable when edges are inserted or removed. Providing theoretic proofs of these observations are currently open and will be the next direction of our future work.




\begin{small}
\noindent\textbf{Acknowledgement} 
The authors would like to thanks the funding agencies NSF DMS-1737812, CNS-1618391, CCF-1535900, DMS-1418255, and AFOSR FA9550-14-1-0193.
\end{small}



%
%
%
\bibliographystyle{splncs04}
\bibliography{ccni,matching,jie,graph_matching,yuylin,publication_gu} 

\newpage
\appendix
\section{Algorithm Pseudo-code}
Algorithm~\ref{algo:ricci_flow} shows the steps of our Ricci flow algorithm. 


\begin{algorithm}
    \SetKwInOut{Input}{Input}
    \SetKwInOut{Output}{Output}

    \Input{An undirected graph $G$.}
    \Output{A weighted graph $G$ in which the Ricci curvature on all edges are the same.}

	Initialize $w_0(x, y)=1$, $\forall \overline{xy}\in E$;

	Compute the Ricci curvature $\kappa_0(x, y)$, $\forall \overline{xy}\in E$;
	
	Update the edge weight by $$w_{i+1}(x,y)\leftarrow w_{i}(x,y)-\epsilon \cdot \kappa_{i}(x,y) \cdot w_{i}(x,y);$$
	
	Normalize the edge weight by $$w_{i+1}(x,y) \leftarrow w_{i+1}(x,y) \cdot \frac{|E|}{\sum_{\overline{xy} \in E}w_{i+1}(x,y)};$$
	
	Repeat $2-4$ until the curvature values do not change much.

    \caption{Ricci flow on a graph}
	\label{algo:ricci_flow}
    
\end{algorithm}

\section{Evaluation Datasets} 
\label{sub:data_sets}

For generality, the network alignment experiments are evaluated over both model and real world network graphs. Here we give an overview of these graphs. The statistics of these graphs are listed in Table~\ref{table:model}. 

\begin{table}[htp]
\begin{small}
\begin{center}
\caption{\small Graph properties for networks.}
\begin{tabular}{|c|c|c|c|c|c|}\hline
Data Set & \#Node & \#Edge  & Avg.Deg. & Diam. & Mean SP \\
\hline
\hline G(n,p)           & 1000    & 4979   & 9.958 & 6 & 3.259 \\
\hline Kleinberg        & 900   & 5462   & 12.138 & 5 & 3.272\\
\hline Pref. Attach.    & 1000  & 3984    & 7.968 & 5 & 3.138\\
\hline Random Regular   & 1000 & 6000    & 12 & 4 & 3.05\\
\hline Random Regular   & 10000 & 40000    & 8 & 7 & 4.816\\
\hline
\hline AS: 3967         & 895    & 2071  & 4.628 & 13 & 5.944\\
\hline Email            & 1133 & 5451  & 9.62 & 8 & 3.606\\
\hline Protein Network  & 2217 & 6418   & 5.789 & 10 & 3.844\\
\hline
\end{tabular}
\label{table:model}
\end{center}
\end{small}
\end{table}

\subsection{Model Networks} 

\paragraph{Erd\H{o}s-R\'{e}nyi} The Erd\H{o}s-R\'{e}nyi model~\cite{erdos59er}, also known as the $G(n, p)$ model, connects each pair of nodes $\{i, j\}$ by an edge with probability $p$ independent of every other edge. We choose $(n,p)=(1000,0.01)$ in our model.

\paragraph{Kleinberg} The Kleinberg graph~\cite{Kleinberg:2000ug} begins with an $n\times n$ squared graph. Each node is represented by $\{(i, j):i,j\in\{1, 2,\cdots, n\}\}$. The distance between two nodes $(i,j)$ and $(k,l)$ is defined by $d((i, j),(k, l)) = |k-i| + |l-j|.$ For any node $u$, $q$ long-range connections to endpoint $v$ are constructed by probability proportional to $d(u,v)^{-r}$. Here we take $(q,r)=(4,2)$.

\paragraph{Preferential Attachment} We choose the Barab\'{a}si-Albert model~\cite{barabasi99emergence} as the preferential attachment model. The model starts with $k$ nodes and then adds nodes one by one. Each newcomer node connects $k$ edges to the existing nodes. For each edge, the endpoint is selected with probability proportional to its current degree. We take $(n,k)=(1000,4)$ in our example.

\paragraph{Random Regular} A random $d-$regular graph is a graph with $n$ nodes and each node randomly connected to other $d$ nodes. We take $(n,d)=(1000,12)$ in our example. 


\subsection{Real Networks} 
\label{sub:datasets_of_real_network_models}

\paragraph{Router Network} A router network graph is composed of routers as nodes and the physical connections as edges. We take the data set `AS:3967, Exodus' from the Rocketfuel project~\cite{Spring2002Rocketfuel}.

\paragraph{Email Network} We take the email communication network at the University Rovira i Virgili. In the graph, nodes are users and edges represent emails between users~\cite{Guimera:2003dm,Kunegis:2013ih}.

\paragraph{Protein Network} We choose the protein-protein interaction network in human cells. Here every node in graph is a protein and an edge represents the interaction between proteins~\cite{Ewing:2007jq,Kunegis:2013ih}.
    

\section{Comparisons With Other Methods} 
\label{sub:matching_evaluations}

We compared our method with other topology-based methods in literature. The brief introduction of these methods are described as follows. All experiments were performed in Python on a 2012 Retina Macbook Pro with 16G ram and 2.6 GHz Intel Core i7. We used CVXPY+ECOS~\cite{cvxpy} as the Linear Programming solver.

\paragraph{Spectral Embedding} The spectral embedding method~\cite{Luxburg:2007:TSC:1288822.1288832} uses the eigenvectors of the graph Laplacian matrix to embed nodes into the corresponding eigenspace. It has been widely applied to graph matching and data clustering such as image segmentation.

\paragraph{Spring Embedding} Spring embedding is an iterative process first introduced by Eades~\cite{eades1984heuristic}. An advanced algorithm was later developed by Fruchterman and Reingold~\cite{fruchterman1991graph}. A graph can be treated as a spring system such that the nodes with initial positions oscillate until they find their own stable positions. The spring embedding can be regarded as an evolution process, therefore, it is useful to observe the change of the same dataset overtime.

\paragraph{IsoRank} IsoRank was proposed by Singh \etal~\cite{Singh2008-hv} for biological network alignment. A matching score is defined on each pair of nodes between two networks. The matching score interprets not only the matching quality of the pair of node but also that of their neighbors. 


\paragraph{Network Similarity Decomposition (NSD)} NSD~\cite{kollias2012network} is an improvement of IsoRank in terms of running time. It is performed by pre-processing two input graphs separately and decomposing the computation of similarity matrix.

\section{Running Time of ATD and OTD.}
\begin{figure}[htbp]
    \centering
    \includegraphics[width=0.75\columnwidth]{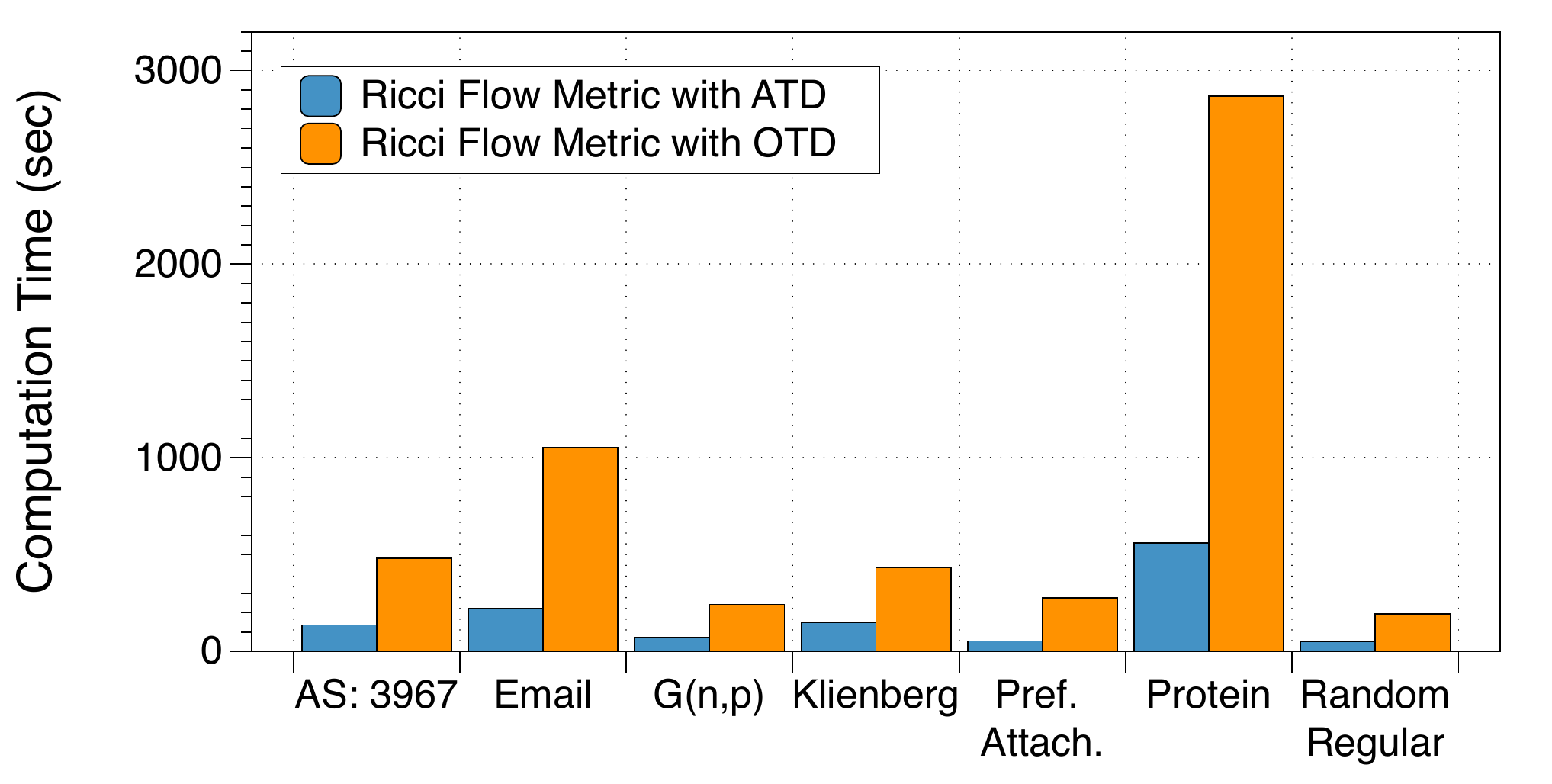}
    \caption{A comparison of computation times between Ricci flow metric with apportioned Ricci curvature and Ollivier Ricci curvature.}
    \label{fig:time-compare}
\end{figure}

Here we analyze the speed performance of Ricci flow metric with ATD and OTD in Figure~\ref{fig:time-compare}. In all settings, the computation time of Ricci flow metric with ATD is up to 3X faster than that of OTD. This is because ATD does not need linear programming computation for computing the optimal transport cost. In fact, most of the computation time in Ricci flow metric by ATD is to compute the all pair shortest path, which can be easily parallelized for further speed improvement.

\section{More Alignment Results}
\label{sec:eval_result}

\begin{figure}[htbp]
	\centering
	\subfigure[G(n, p)]{
		\label{fig:model:subfig:gnp}
		\includegraphics[width=0.48\columnwidth]{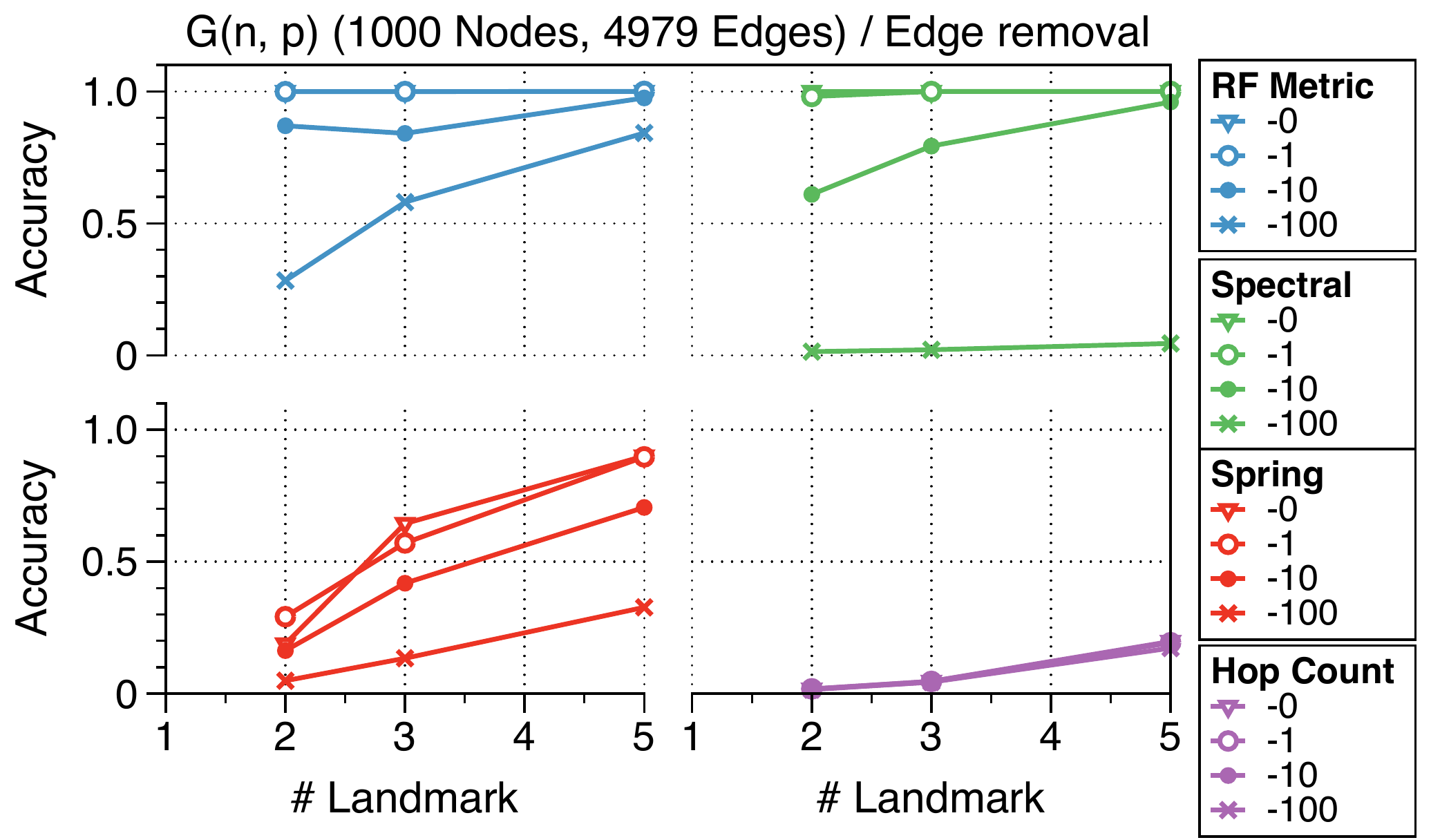}}
	\subfigure[Preferential Attachment]{
		\label{fig:model:subfig:prefAttach}
		\includegraphics[width=0.48\columnwidth]{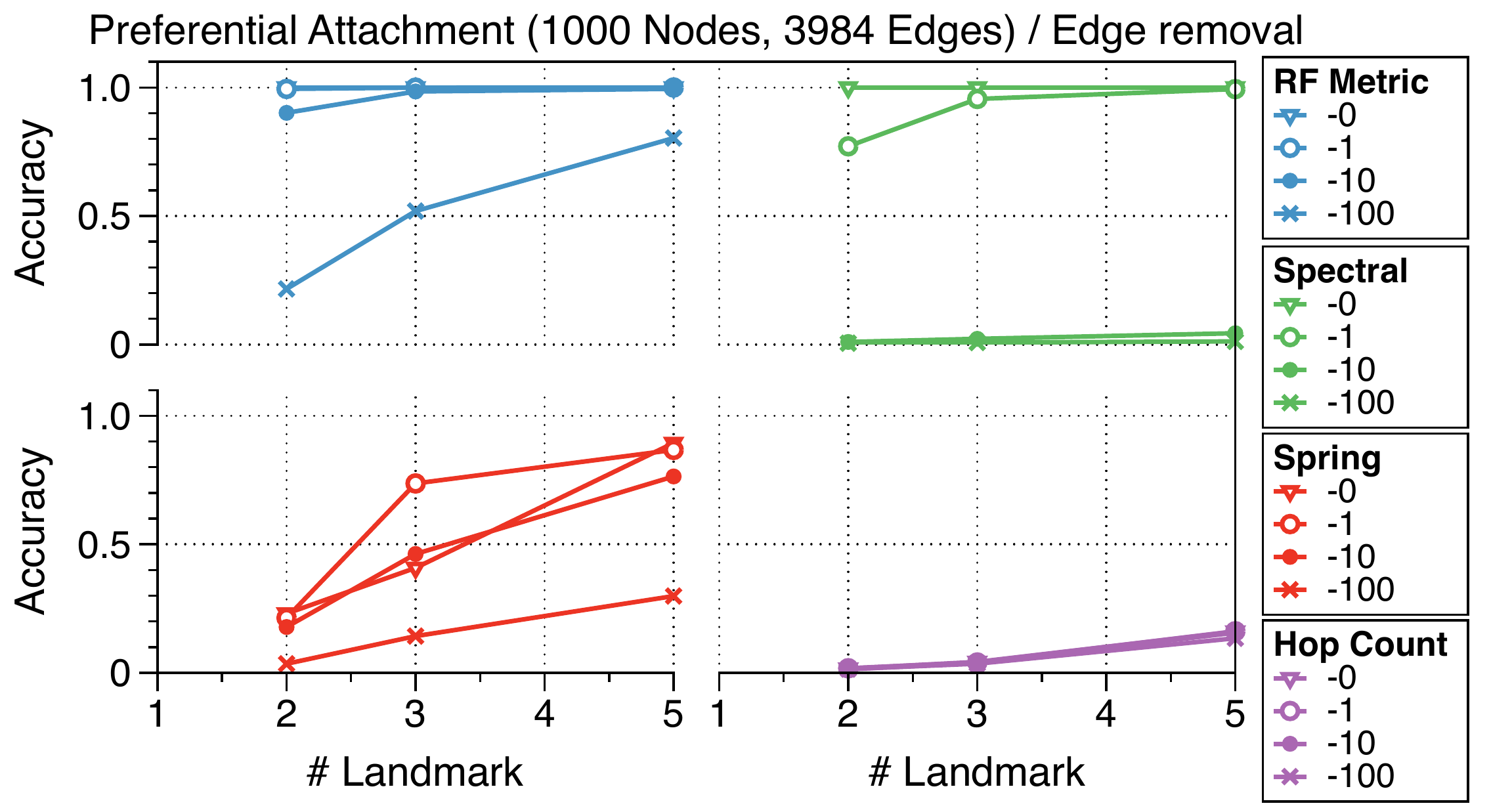}}
	\subfigure[Random Regular]{
		\label{fig:model:subfig:rr}
		\includegraphics[width=0.48\columnwidth]{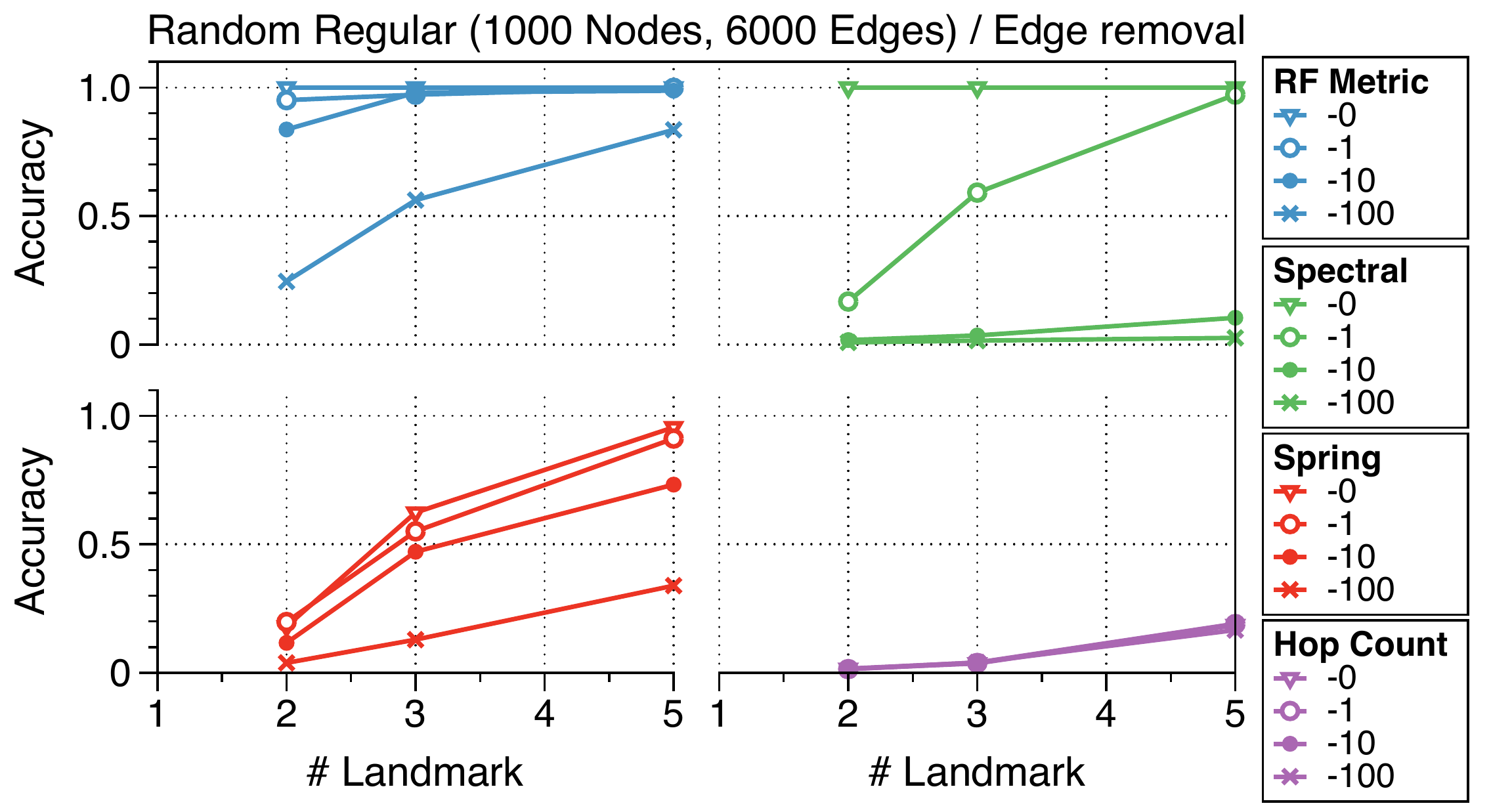}}
	\subfigure[Kleinberg]{
		\label{fig:model:subfig:klienberg}
		\includegraphics[width=0.48\columnwidth]{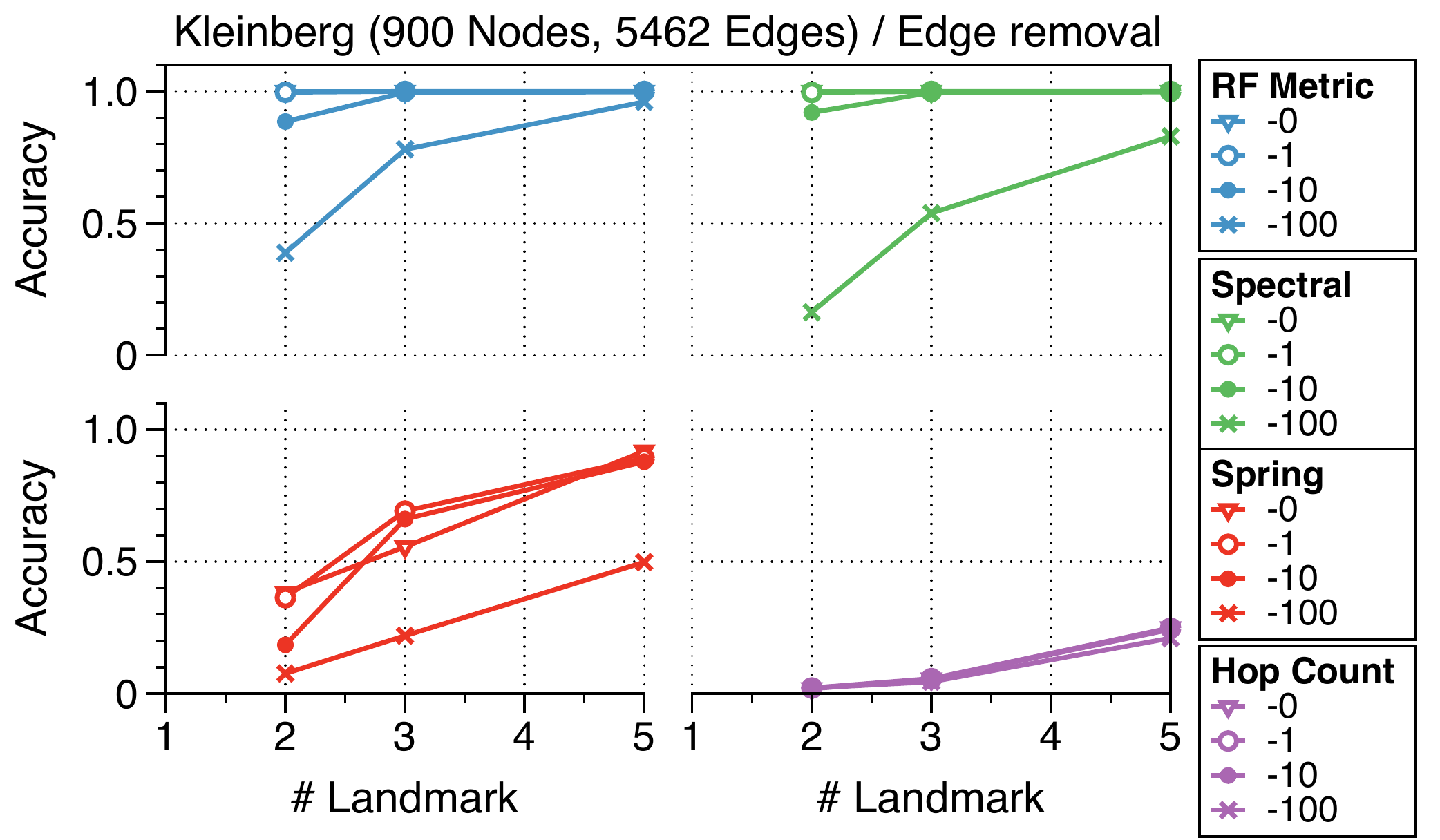}}
    \vspace{-4mm}
	\caption{\scriptsize Result of approximate graph matching on model networks using Hungarian algorithm.}
    \vspace{-4mm}
	\label{fig:model} 
		
\end{figure}

In the following experiment we consider the case with random edge removal, spectral embedding with $2$ dimensions, and up to $5$ landmarks. We use Ricci Flow metric with ATD, Hungarian algorithm, and connected equivalence in the following results. Figure~\ref{fig:model} shows the results of graph matching on different model networks, when $G_2$ removes $0, 1, 10, 100$ edges from $G_1$ respectively. In all the graphs, Ricci flow metric yields the best performance of matching accuracy, even with only $2$ or $3$ landmarks, and with a large number of removed edges. The accuracy of spring embedding and hop count are low when the number of landmarks is small.

Notice that spectral embedding shows sensitivity with edge removal. In a random regular graph, even with only $1$ edge removed, the accuracy drops dramatically. This is because the stability of spectral embedding is highly influenced by the connectivity of the graph. If a graph has more tree-like edges, its spectral embedding is more likely to be affected by edge removal. 

\begin{figure}[htbp]
	\centering
	\subfigure[AS: 3967, Exodus]{
		\label{fig:real:subfig:3967}
		\includegraphics[width=0.48\columnwidth]{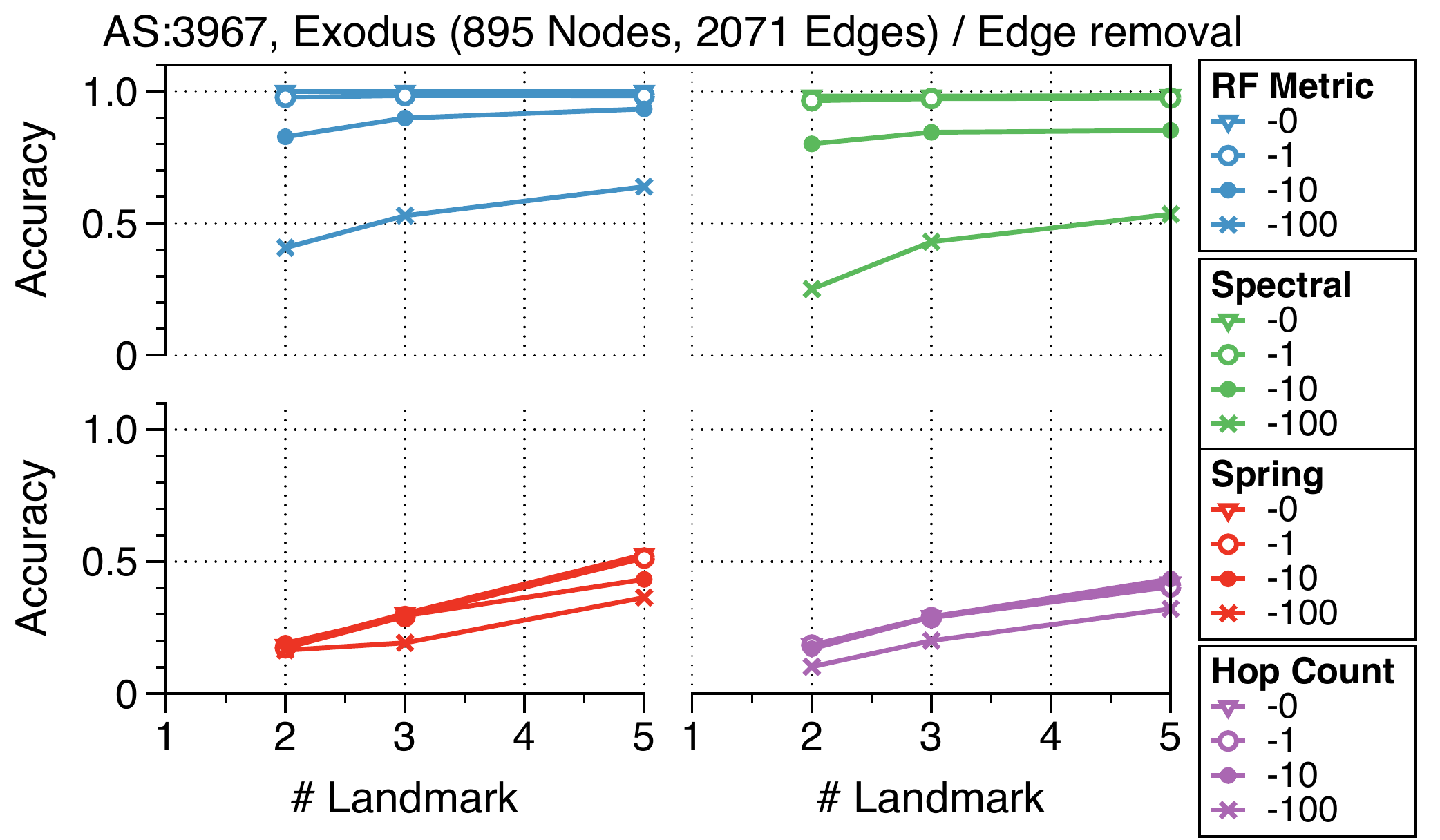}}
	\subfigure[Email]{
		\label{fig:real:subfig:email}
		\includegraphics[width=0.48\columnwidth]{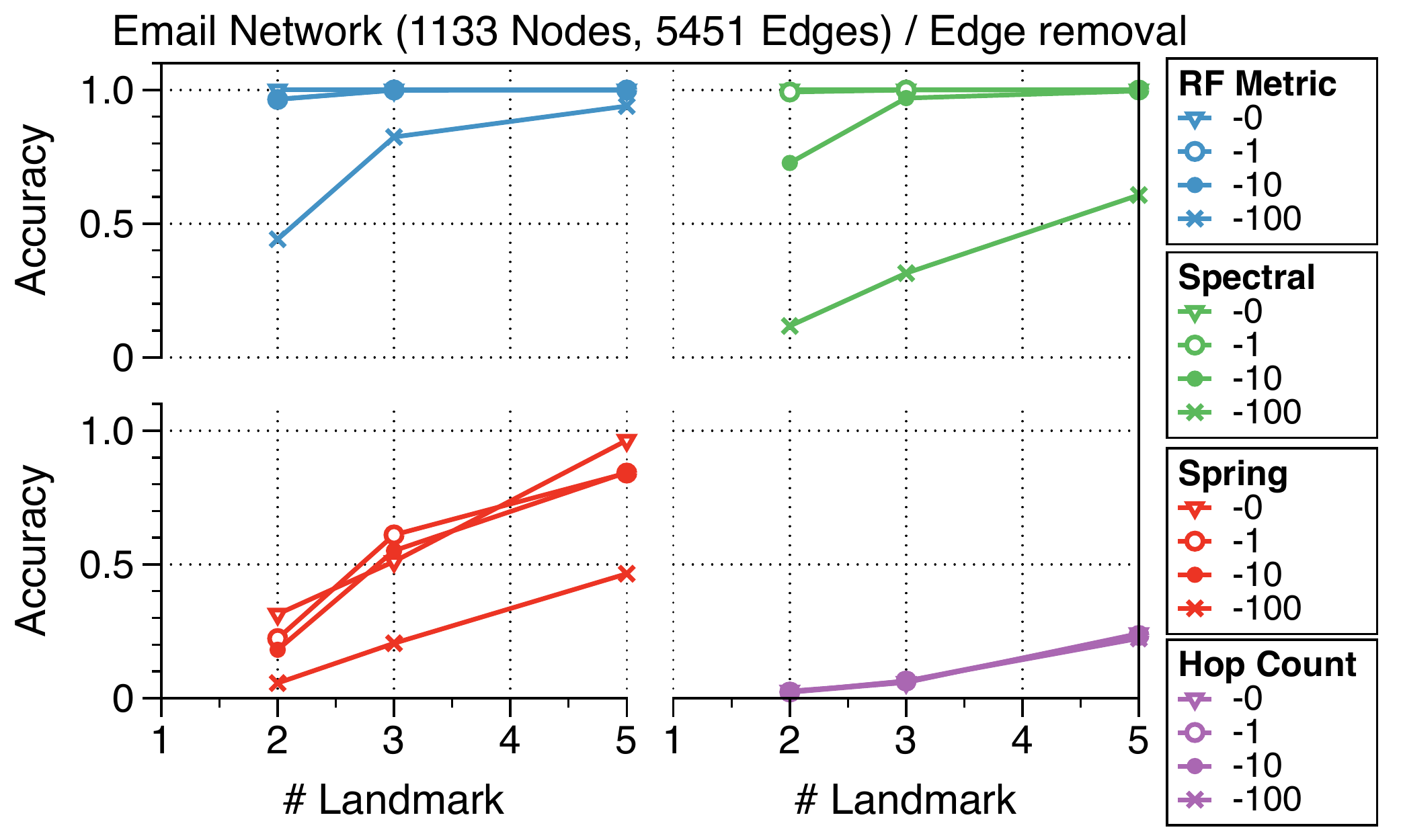}}
	\subfigure[Protein-protein interaction]{
		\label{fig:real:subfig:ppi}
		\includegraphics[width=0.48\columnwidth]{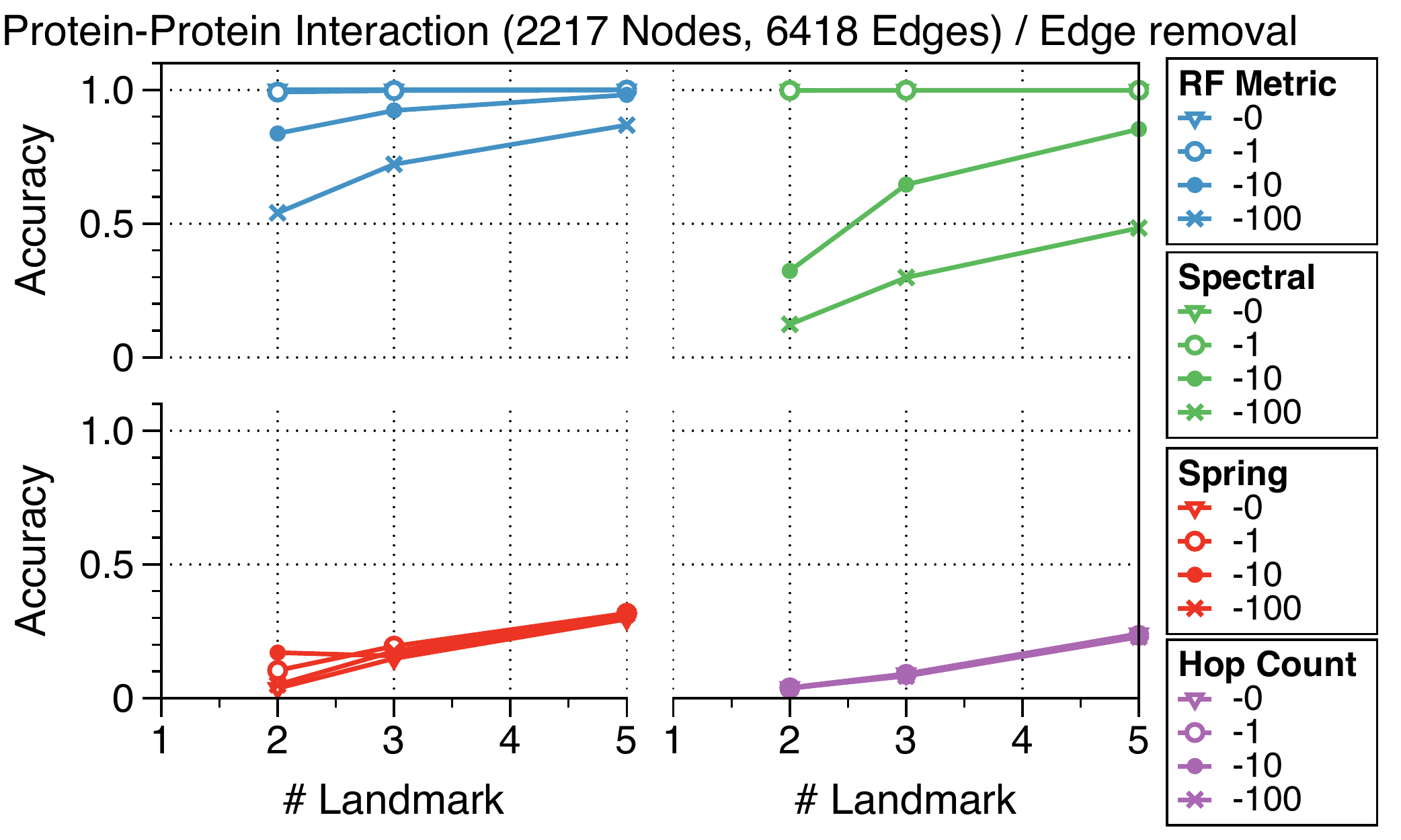}}
    \vspace{-4mm}
	\caption{\scriptsize Result of approximate graph matching on real networks using Hungarian algorithm.}
    
	\label{fig:real} 
		
\end{figure}

Figure~\ref{fig:real} illustrates the matching results on real networks. Again, Ricci flow metric performs the best on all of the graphs. 
Spectral embedding behaves relatively better on real networks than on model networks. This is because real networks do have high clustering coefficient and edges with high Jaccard coefficient (Figure~\ref{fig:jaccard:subfig:real}). Notice that the matching accuracy of spring embedding and hop count on AS:3967 and protein-protein interaction graph are low, this may be caused by the larger diameter of these two graphs.

\begin{figure}[htbp]
	\centering
	\subfigure[AS: Greedy Matching]{
		\label{fig:greedy:subfig:as}
		\includegraphics[width=0.48\columnwidth]{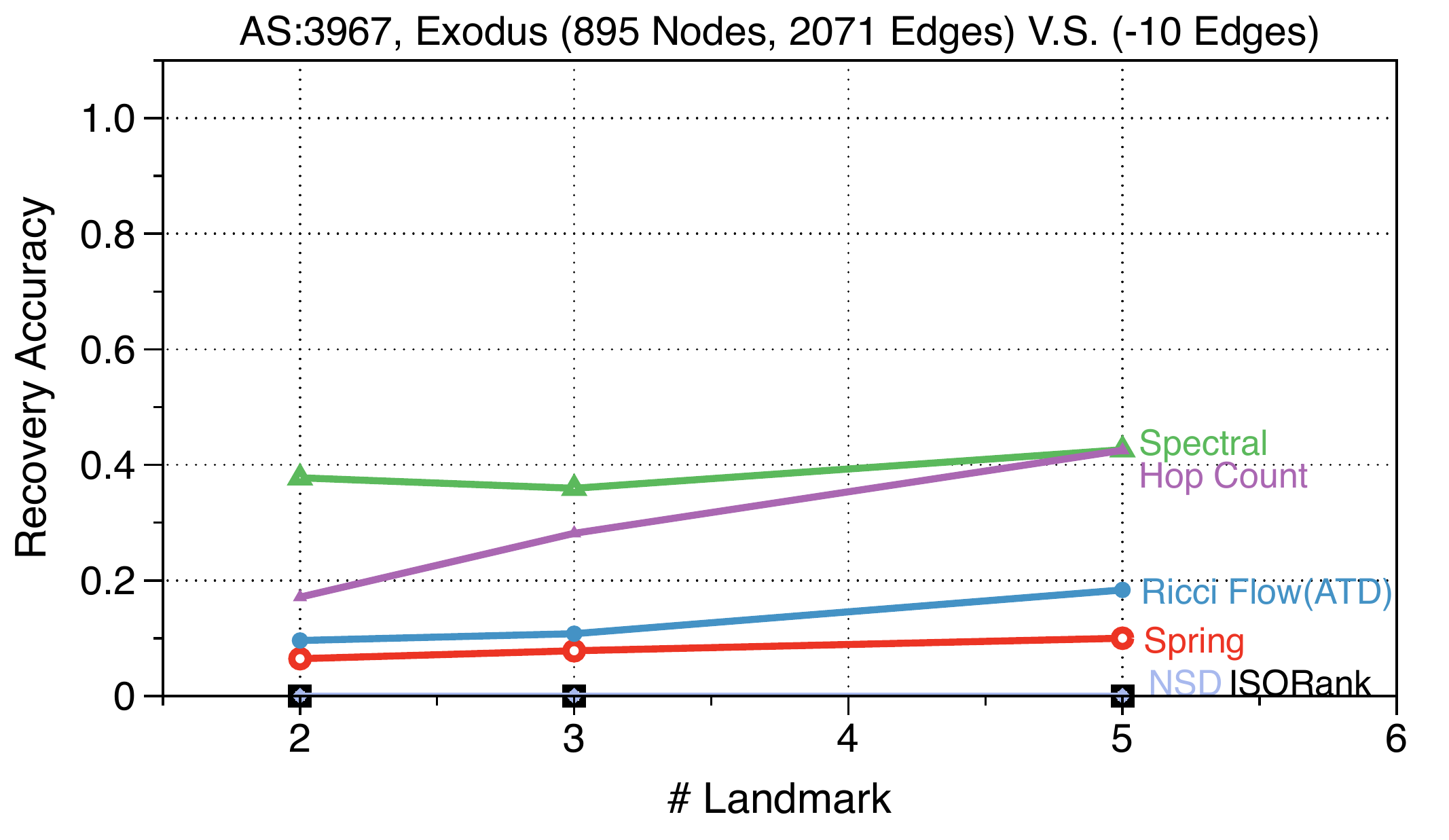}}
	\subfigure[PPI: Greedy Matching]{
		\label{fig:greedy:subfig:ppi}
		\includegraphics[width=0.48\columnwidth]{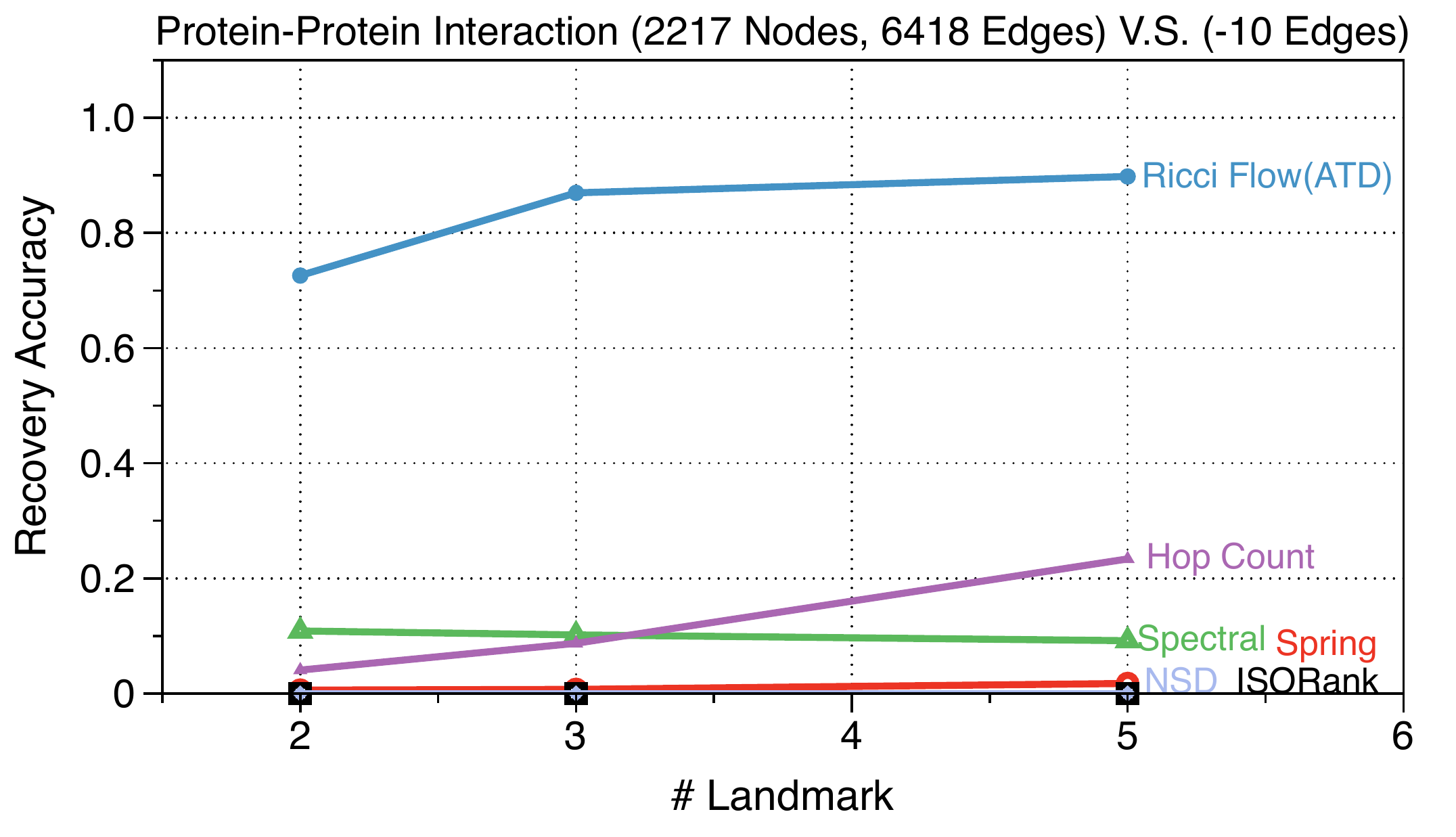}}
	\subfigure[Random Regular: Greedy Matching]{
		\label{fig:greedy:subfig:rr}
		\includegraphics[width=0.48\columnwidth]{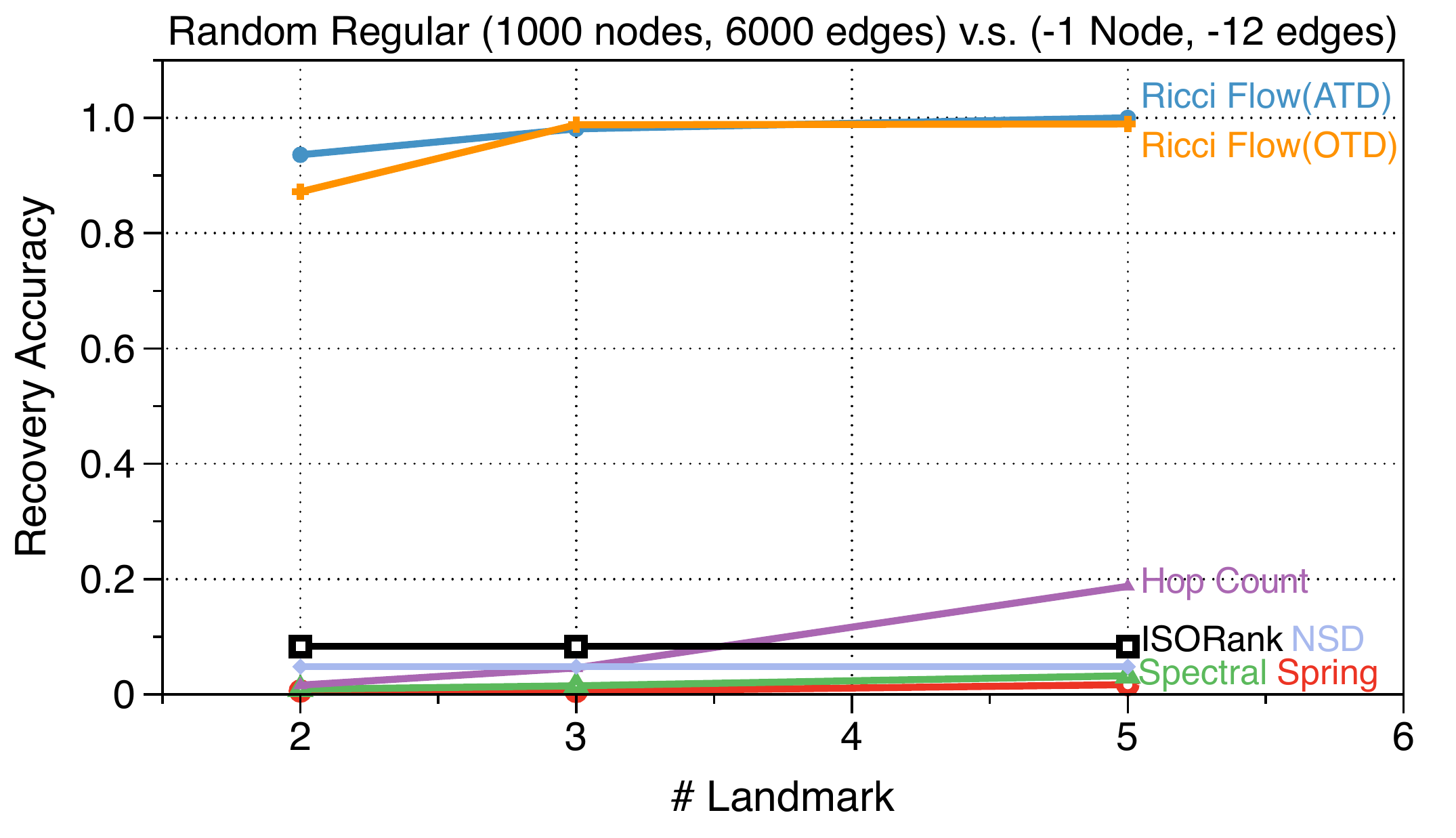}}
    \vspace{-4mm}
	\caption{\scriptsize Result of noisy graph alignment with greedy matching and connected equivalence.}
    \vspace{-4mm}
	\label{fig:greedy} 
		
\end{figure}

In Figure~\ref{fig:greedy}, the result of noisy graph alignment using greedy matching instead of Hungarian algorithm is presented. With greedy matching, most of the methods experience performance degradation except for Ricci flow metric. In AS graph (Figure~\ref{fig:greedy:subfig:as}), all methods performs poorly with greedy matching because of the graph's tree like structure and large graph diameter.

%
%
%
%

\section{Tree-Like Graphs} 
\label{sec:jaccard_coefficient}

\begin{figure}[htbp]
    \centering
    \subfigure[Model Networks]{
        \vspace{-4mm}
        \label{fig:jaccard:subfig:model}
        \includegraphics[width=0.48\columnwidth]{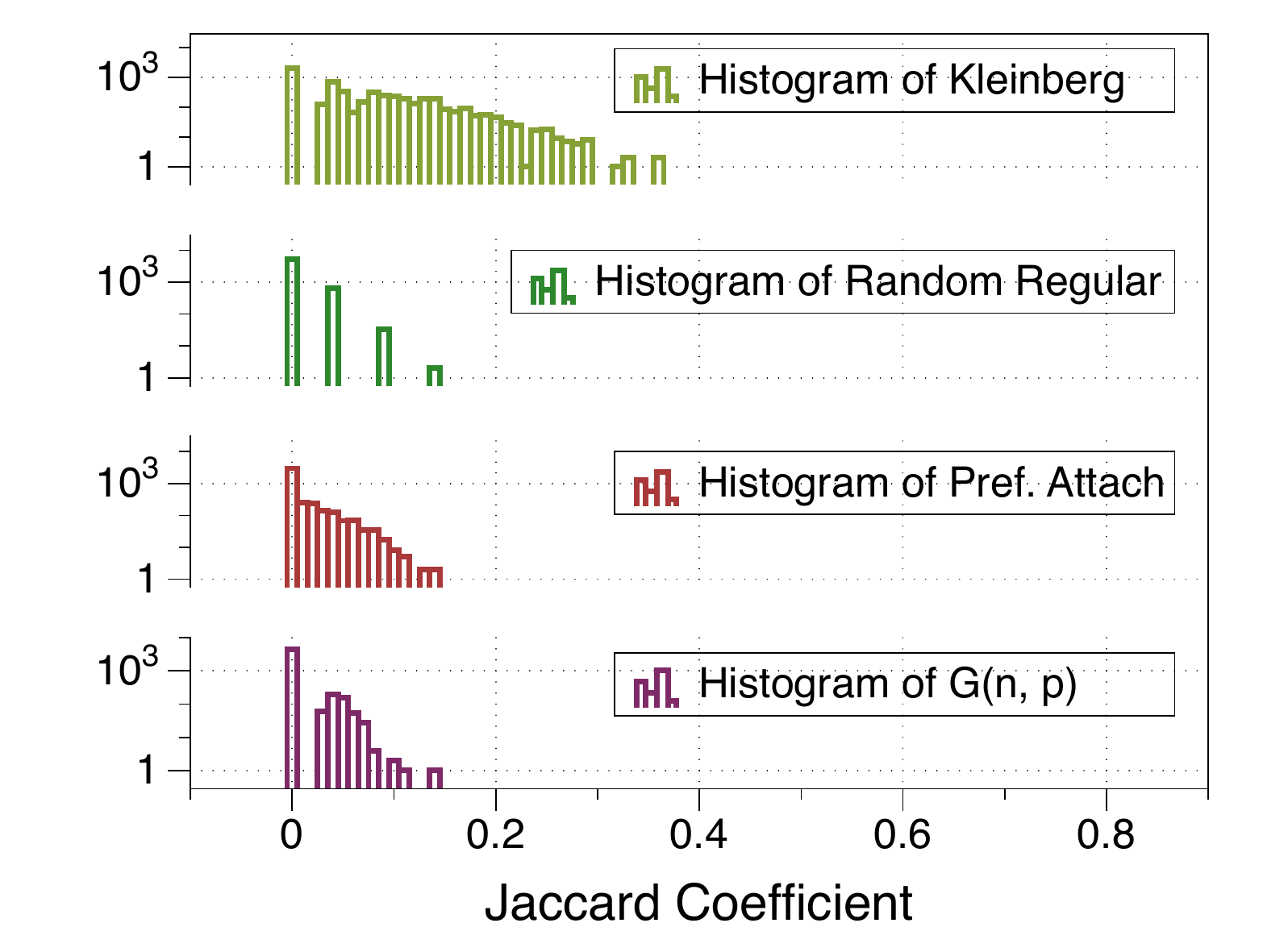}}
    \subfigure[Real Networks]{
        \label{fig:jaccard:subfig:real}
        \includegraphics[width=0.48\columnwidth]{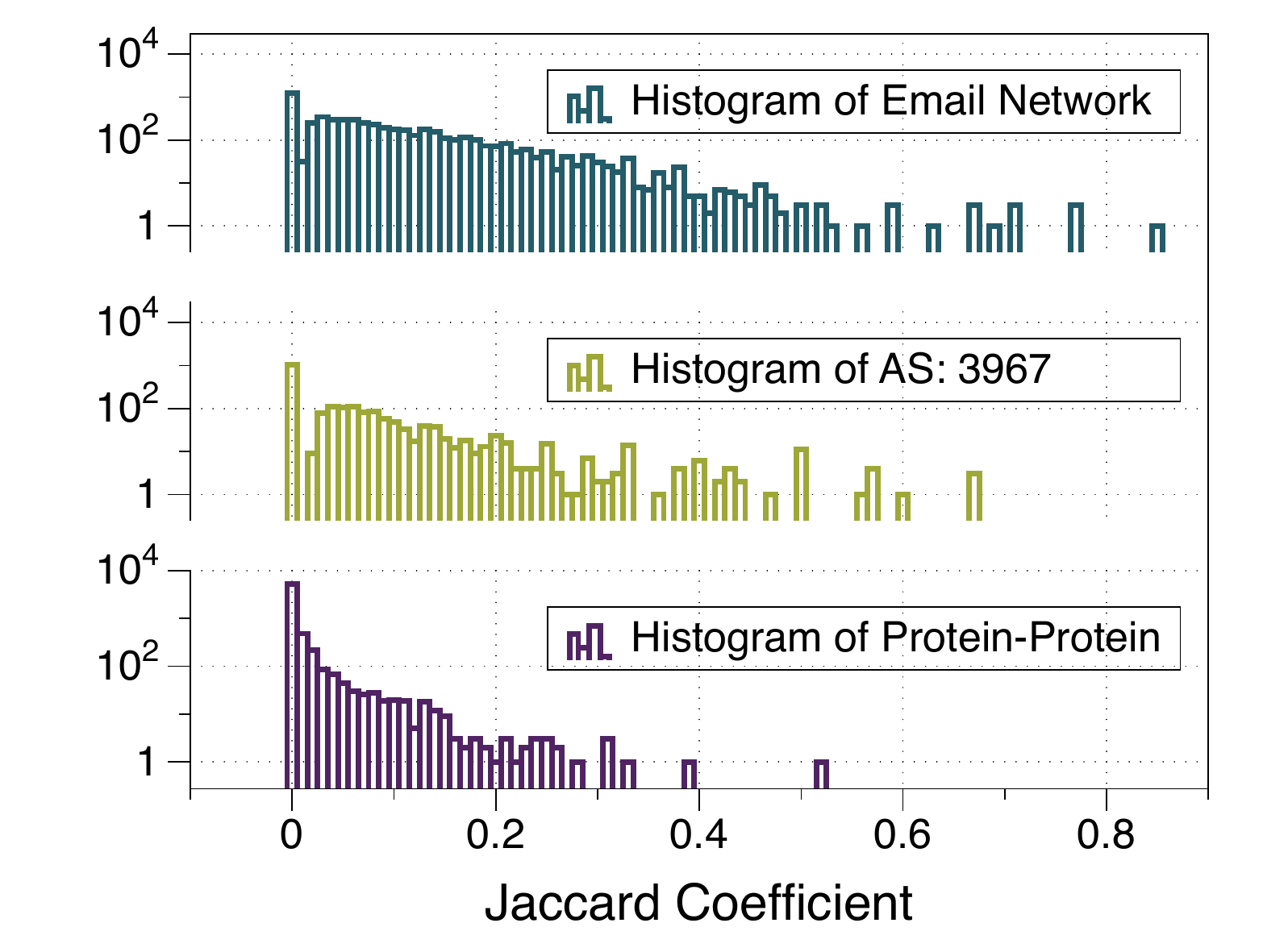}}
    \vspace{-4mm}
    \caption{\scriptsize Histogram of Jaccard Coefficient on edge over networks.}
    \vspace{-4mm}
    \label{fig:jaccard} 

\end{figure}

To capture how much would the results be affected by tree-like edges in the graphs, we evaluate the Jaccard coefficient on each edge as displayed in Figure~\ref{fig:jaccard:subfig:model}. For an edge $\overline{uv}$, the Jaccard coefficient of this edge is defined as $\pi_u\cap\pi_v/\pi_u\cup\pi_v$. If an edge has a large Jaccard coefficient, it is likely to be an interior edge of a cluster or a community; if an edge has a small Jaccard coefficient, it then acts more like a tree edge or a bridge between clusters. In Figure~\ref{fig:jaccard:subfig:model}, it is clear that Kleinberg model has more edges with high Jaccard coefficients, its spectral embedding is more robust and supports our results in Figure~\ref{fig:model:subfig:klienberg}.

\section{More Tests of Spectral and Spring Embedding}
\label{sec:spectral-dim}

\begin{figure}[htbp]
    \centering
    \vspace{-4mm}
       \includegraphics[width=0.6\columnwidth]{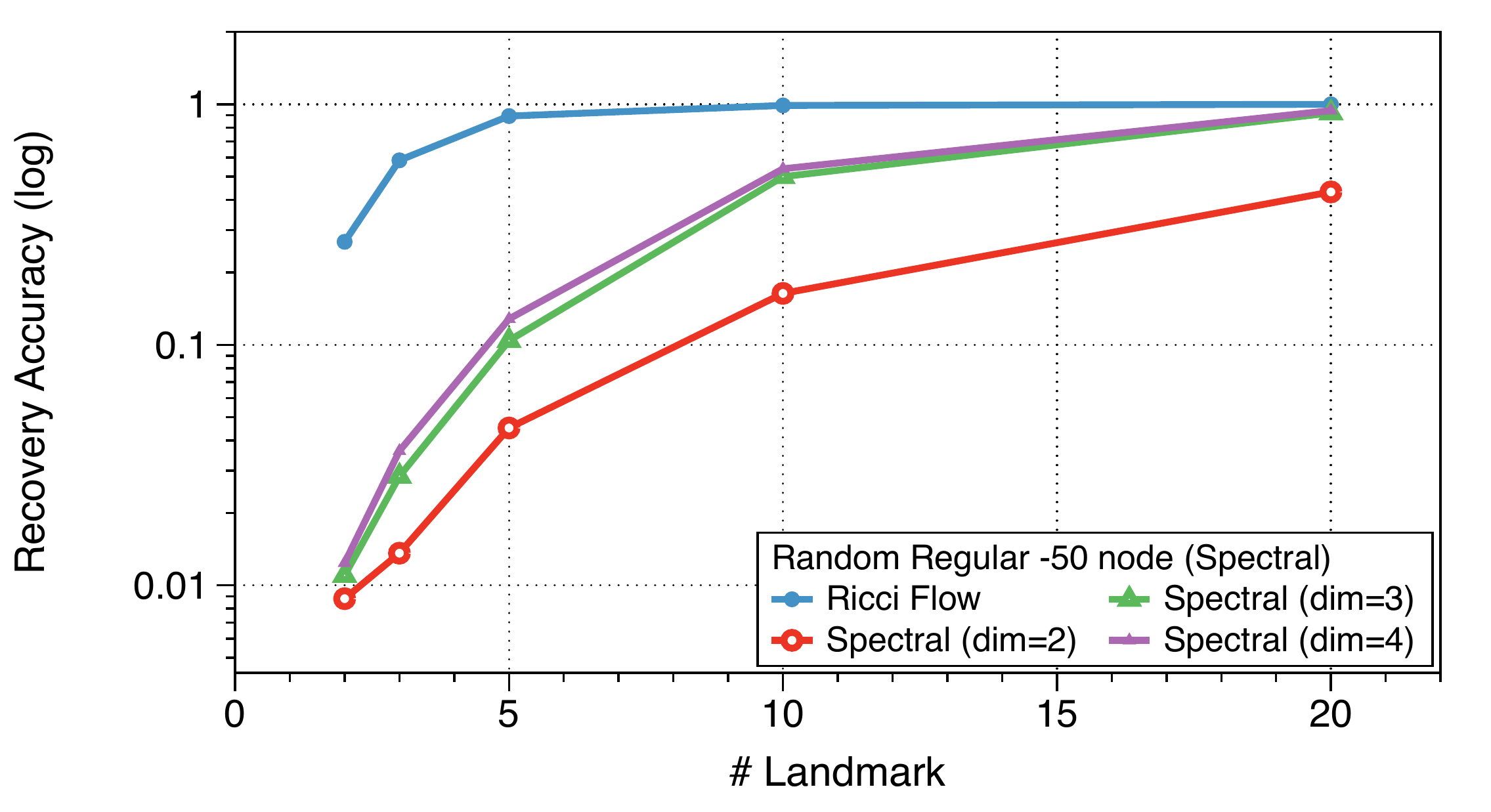}
    \vspace{-4mm}
    \caption{\scriptsize A comparison of graph alignment results between Ricci flow metric and spectral embedding of different dimensions.}
    \vspace{-4mm}
    \label{fig:spectral}
\end{figure}

In Figure~\ref{fig:spectral}, we tested the influence of spectral embedding into Euclidean space in Euclidean spaces of different dimensions. The improvement of alignment accuracy with increasing dimensionality is still limited. 

\end{document}